\documentclass{nature}
\usepackage{graphicx}
\usepackage{blindtext}
\usepackage{amsmath}
\usepackage{mathtools}
\usepackage{multirow}
\usepackage{pdflscape}
\usepackage{ragged2e}
\usepackage{lineno}
\usepackage[hidelinks]{hyperref}
\usepackage{color}

\makeatletter
\let\saved@includegraphics\includegraphics
\AtBeginDocument{\let\includegraphics\saved@includegraphics}
\renewenvironment*{figure}{\@float{figure}}{\end@float}
\makeatother

\newcommand{\hst}{\mbox{HST}}
\newcommand{\hstlong}{\mbox{\textit{Hubble Space Telescope}}}

\newcommand{\jwstlong}{\mbox{\textit{James Webb Space Telescope}}}
\newcommand{\ariel}{\mbox{ARIEL}}
\newcommand{\wfc}{\mbox{WFC3}}

\newcommand{\grism}{\mbox{G141}}
\newcommand{\planet}{\mbox{K2-18\,b}}

\newcommand{\taurex}{\mbox{TauREx}}

\newcommand{\phoenix}{\mbox{PHOENIX}}
\newcommand{\iraclis}{\mbox{\textit{Iraclis}}}
\newcommand{\plc}{\mbox{\textit{PyLightcurve}}}

\newcommand\aj{{Astron. J.}}
\newcommand\apj{{Astrophys. J.}}
\newcommand\apjl{{Astrophys. J. Lett.}}     
\newcommand\aap{{Astron. Astrophys.}}
\newcommand\mnras{{Mon. Not. R. Astron. Soc.}}
\newcommand\nat{{Nature}}
\newcommand\jqsrt{{J. of Quant. Spec. \& Rad. Transf.}}
\newcommand\pasp{{Publ. Astron. Soc. Pac.}}

\begin{document}

\begin{center}
\textbf{\Large{Water vapour in the atmosphere of the habitable-zone eight Earth-mass planet K2-18 b}}
\end{center}

\noindent Angelos Tsiaras$^{1,*}$, Ingo\,P. Waldmann$^{1}$, Giovanna Tinetti$^{1}$, Jonathan 
Tennyson$^{1}$ \& Sergey N. Yurchenko$^{1}$

\noindent {\small{ \textit{$^1$Department of Physics \& Astronomy, University 
College London, Gower Street, WC1E 6BT London, UK \\
$^*$Corresponding author}}}

\justify

\textbf{In the past decade, observations from space and ground have found H$_2$O to be the most abundant molecular species, after hydrogen, in the atmospheres of hot, gaseous, extrasolar planets\cite{Tinetti2007B2007ApJ...654L..99T,Grillmair2008,Fraine2014,Macintosh2015,Tsiaras2018}. Being the main molecular carrier of oxygen, H$_2$O is a tracer of the origin and the evolution mechanisms of planets. For temperate, terrestrial planets, the presence of H$_2$O is of great significance as an indicator of habitable conditions. Being small and relatively cold, these planets and their atmospheres are the most challenging to observe, and therefore no atmospheric spectral signatures have so far been detected\cite{deWit2018}. Super-Earths -- planets lighter than ten M$_\oplus$ -- around later-type stars may provide our first opportunity to study spectroscopically the characteristics of such planets, as they are best suited for transit observations. Here we report the detection of an H$_2$O spectroscopic signature in the atmosphere of \planet\ -- an eight M$_\oplus$ planet in the habitable-zone of an M-dwarf\cite{Montet2015} -- with high statistical confidence (ADI\cite{Tsiaras2018} = 5.0, $\sim$3.6$\sigma$\cite{Benneke2013, Waldmann2015B2015ApJ...802..107W}). In addition, the derived mean molecular weight suggests an atmosphere still containing some hydrogen. The observations were recorded with the Hubble Space Telescope/WFC3 camera, and analysed with our dedicated, publicly available, algorithms\cite{Tsiaras2018,Waldmann2015B2015ApJ...802..107W}. While the suitability of M-dwarfs to host habitable worlds is still under discussion\cite{Segura2005,Wordsworth2011,Leconte2013,Turbet2016}, \planet\ offers an unprecedented opportunity to get insight into the composition and climate of habitable-zone planets.}

Atmospheric characterisation of super-Earths is currently within reach of the Wide Field Camera 3 (\wfc) onboard the \hstlong\ (\hst), combined with the recently implemented spatial scanning observational strategy\cite{Deming2013}. The spectra of three hot transiting planets with radii less than 3.0 R$_{\oplus}$ have been published so far: GJ-1214\,b\cite{Kreidberg2014B2014Natur.505...69K}, HD\,97658\,b\cite{Knutson2014B2014ApJ...794..155K} and 55\,Cnc\,e\cite{Tsiaras2016B2016ApJ...820...99T}. The first two do not show any evident transit depth modulation with wavelength, suggesting an atmosphere covered by thick clouds or made of molecular species heavier than hydrogen, while only the spectrum of 55\,Cnc\,e has revealed a light-weighted atmosphere, suggesting H/He still being present. In addition, transit observations of six temperate Earth-size planets around the ultra-cool dwarf TRAPPIST-1 -- planets b, c, d, e, f\cite{deWit2018}, and g\cite{Wakeford2019} -- have not shown any molecular signatures and have excluded the presence of cloud-free, H/He atmospheres around them.

\planet\ was discovered in 2015 by the \textit{Kepler} spacecraft\cite{Montet2015}, and it is orbiting around an M2.5 ($\mathrm{[Fe/H]} = 0.123 \pm 0.157 \, \mathrm{dex}$,  $T_\mathrm{eff} = 3457 \pm 39 \, \mathrm{K}$, $M_* = 0.359 \pm 0.047 \, M_{\odot}$, $R_* = 0.411 \pm 0.038 \, R_{\odot}$)\cite{Benneke2017} dwarf star, 34\,pc away from the Earth. The star-planet distance of 0.1429\,AU\cite{Benneke2017} suggests a planet within the star's habitable zone ($\sim$\,0.12\,--\,0.25\,AU)\cite{Kopparapu2013}, with effective temperature between 200\,K and 320\,K, depending on the albedo and the emissivity of its surface and/or its atmosphere. This crude estimate accounts for neither possible tidal energy sources\cite{Valencia2018} nor atmospheric heat redistribution\cite{Wordsworth2011, Turbet2016}, which might be relevant for this planet. Measurements of the mass and the radius of \planet\ ($M_\mathrm{p} = \,7.96 \pm 1.91 \, M_\oplus$\cite{Cloutier2017},  $R_\mathrm{p} = 2.279 \pm 0.0026\, R_\oplus$\cite{Benneke2017}) yield a bulk density of 3.3$\pm$1.2 g/cm\cite{Cloutier2017}, suggesting either a silicate planet with an extended atmosphere around  or an interior composition with an H$_2$O mass fraction lower that 50\%\cite{Valencia2013,Zeng2016,Cloutier2017}. 

We analyse here eight transits of \planet\, obtained with the WFC3 camera onboard the Hubble Space Telescope. We used our specialised, publicly available, tools\cite{Tsiaras2018, Waldmann2015B2015ApJ...802..107W} to perform the end-to-end analysis from the raw \hst\ data to the atmospheric parameters. The accuracy of the techniques used here have been demonstrated through the largest consistently analysed catalogue of exoplanetary spectra from \wfc \cite{Tsiaras2018}. Details on the data analysis can be found in the Methods section. Also, links to the data and the codes used can be found in the Data availability and Code availability sections, respectively. Alongside with the data we provide descriptions of the data structures and instructions on how to reproduce the results presented here. 
Our analysis resulted in the detection of an atmosphere around K2-18 b with an ADI\cite{Tsiaras2018} (a positively defined logarithmic Bayes Factor) of 5.0, or approximately 3.6$\sigma$ confidence\cite{Benneke2013, Waldmann2015B2015ApJ...802..107W}, making K2-18 b the first habitable-zone planet in the super-Earth mass regime (1-10\,M$_\oplus$) with an observed atmosphere around it. 

More specifically, nine transits of \planet\ were observed as part of the \hst\ proposals 13665 and 14682 (PI: Bj{\"o}rn Benneke) and the data are available through the MAST Archive (see Data Availability section). Each transit was observed during five \hst\ orbits, with the \grism\ infrared grism (1.1 - 1.7\,$\mu$m), and each exposure was the result of 16 up-the-ramp samples in the spatially scanning mode. The ninth transit observation suffered from pointing instabilities and therefore we decided not to include it in this analysis. We extracted the white and the spectral light curves from the reduced images, following our dedicated methodology\cite{Tsiaras2016B2016ApJ...832..202T, Tsiaras2016B2016ApJ...820...99T, Tsiaras2018}, which has been integrated into an automated, self-consistent, and user-friendly Python package named \iraclis\ (see Code Availability section). No systematic variations of the white light-curve $R_\mathrm{p}/R_*$ appeared between the eight different observations. This level of stability among the extracted broad-band transit depths is not always guaranteed, as consistency problems among different observations emerged in previous analyses\cite{Knutson2014B2014ApJ...794..155K, Tsiaras2018}.

In our analysis, we found that the measured mid-transit times where not consistent with the expected ephemeris\cite{Benneke2017}. We used these results to refine the ephemeris of K2-18\,b to be: $P=32.94007\pm0.00003$\,days and $T_0=2457363.2109\pm0.0004$ BJD$_\mathrm{TDB}$\cite{Eastman2010}, where $P$ is the period, $T_0$ is the mid-time of the transit. However, the ephemeris calculated only from the HST data is not consistent with the original detection of K2-18\,b. One possibility is that the very sparse data from K2 are not sufficient to give a confident result. Another possibility is that we observe significant transit time variations (TTVs) caused by the other planet in the system, K2-18\,c\cite{Cloutier2017}, but more observations over a long period of time are necessary to disentangle the two scenarios. In addition, we used the detrended and time-aligned -- i.e. with TTVs removed -- white light curves to also refine the orbital parameters and found them to be: $a/R_*=81.3\pm1.5$ and $i=89.56\pm0.02$\,deg, where $a/R_*$ is the orbital semi-major axis normalised to the stellar radius, and $i$ is the orbital inclination.

We extracted eight transmission spectra of \planet\ and combined them, using a weighted average, to produce the final spectrum (Table \ref{tab:spectrum}). We interpreted the planetary spectrum using our spectral retrieval algorithm \taurex \cite{Waldmann2015B2015ApJ...813...13W, Waldmann2015B2015ApJ...802..107W} (see Code Availability section) which combines highly accurate line lists\cite{Tennyson2016} (see Data Availability section) and Bayesian analysis. At an initial stage, we modelled the atmosphere of \planet\ including all potential absorbers in the observed wavelength range -- i.e. H$_2$O, CO, CO$_2$, CH$_4$ and NH$_3$. However, we found that only the spectroscopic signature of water vapour is detected with high-confidence, so we continued our analysis only with this molecule as trace-gas. We modelled the atmosphere following three approaches:
\begin{itemize} 
\item a cloud-free atmosphere containing only H$_2$O and H$_2$/He
\item a cloud-free atmosphere containing H$_2$O, H$_2$/He and N$_2$ (N$_2$ acted as proxy for ``invisible" molecules not detectable in the WFC3 bandpass but contributing to the mean molecular weight), and
\item a cloudy (flat-line model) atmosphere containing only H$_2$O and H$_2$/He.
\end{itemize} 

We retrieved a statistically significant atmosphere around \planet\ in all simulations (Figure \ref{fig:retrievals}), and assessed the strength of the detection using the Atmospheric Detectability Index\cite{Tsiaras2018} (ADI), which represents the positively defined logarithmic Bayes Factor, where the null hypothesis is a model that  contains no active trace gases, Rayleigh scattering or collision induced absorption -- i.e. a flat spectrum. The retrieval simulations yield an atmospheric detection with an ADI of 5.0, 4.7 and 4.0, respectively. Such ADIs correspond to approximately a 3.6, 3.5, and 3.3\,$\sigma$ detection\cite{Benneke2013, Waldmann2015B2015ApJ...802..107W}, respectively. This marks the first atmosphere detected around a habitable-zone super-Earth with such a high level of confidence. While the H$_2$O + H$_2$/He case appears to be the most favourable, this preference is not statistically significant. 

As far as the composition is concerned, retrieval models confirm the presence of water vapour in the atmosphere of \planet\ in all the cases studied with high statistical significance. However, it is not possible to constrain either its abundance or the mean molecular weight of the atmosphere. For the H$_2$O + H$_2$/He case, we found the abundance of H$_2$O to be between 50\% and 20\%, while for the other two cases between 0.01\% and 12.5\%. The atmospheric mean molecular weight can be between 5.8 and 11.5 amu in the H$_2$O + H$_2$/He case, and between 2.3 and 7.8 amu for the other cases. These results indicate that a non-negligible fraction of the atmosphere is still made of H/He. Additional trace-gases -- e.g. CH$_4$, NH$_3$ -- cannot be excluded, despite not being identified with the current observations: the limiting S/N and wavelength coverage of HST/WFC3 do not allow the detection of other molecules. 

The results presented here confirm the existence of a detectable atmosphere around \planet, making it one of the most interesting known targets for further atmospheric characterisation with future observatories, like the \jwstlong\ (0.6\,$\mu$m and 28\,$\mu$m) and the European Space Agency \ariel\ mission\cite{Tinetti2018} (0.5\,$\mu$m and 7.8\,$\mu$m). The wider wavelength coverage of these instruments will provide information on the presence of additional molecular species and on the temperature-pressure profile of the planet, towards studying the planetary climate and potential habitability. While the subject of habitability for temperate planets around late-type stars is a subject of active discussion\cite{Segura2005,Wordsworth2011,Leconte2013,Turbet2016} and real progress requires significantly improved observational constraints, the analysis presented here provides the first direct observation of a molecular signature from a habitable-zone exoplanet, connecting these theoretical studies to observations.



\paragraph*{Acknowledgements} This project has received funding from the European Research Council (ERC) under the European Union's Horizon 2020 research and innovation programme (grant agreements 758892, ExoAI;  776403/ExoplANETS A) and under the European Union's Seventh Framework Programme (FP7/2007-2013)/ ERC grant agreement numbers 617119 (ExoLights) and 267219 (ExoMol). We furthermore acknowledge funding by the Science and Technology Funding Council (STFC) grants: ST/K502406/1 and ST/P000282/1. The data used here were obtained by the \hstlong\ as part of the 13665 and 14682 GO proposals (PI: Bj{\"o}rn Benneke).

\paragraph*{Author Contributions} A.T. performed the data analysis and developed the HST analysis software \iraclis; I.P.W developed the atmospheric retrieval software \taurex; G.T. contributed to the interpretation of the results; J.T. and S.N.Y. coordinate the ExoMol project. All authors discussed the results and commented on the manuscript.

\paragraph*{Financial Interests} The authors declare no competing financial interests.

\paragraph*{Materials \& Correspondence} Correspondence and requests for materials should be addressed to A.T. (atsiaras@star.ucl.ac.uk) or I.P.W. (ingo@star.ucl.ac.uk).

\begin{figure}
	\centering
	\includegraphics[width=\textwidth]{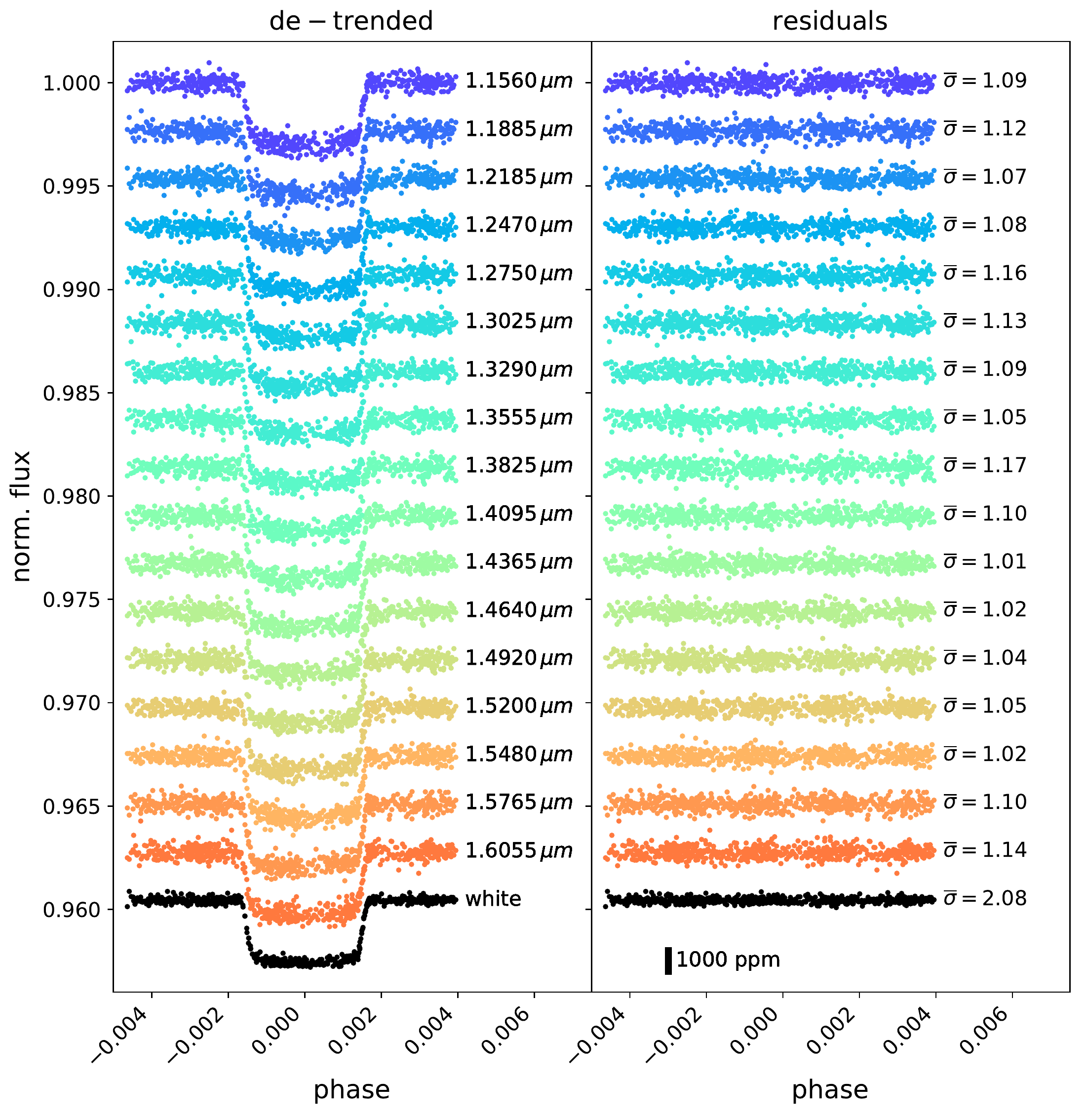}
	\caption{Analysis of the \planet\ white (black points) and spectral 
(coloured points) light curves, plotted with an offset for clarity. Left: Overploted detrended light curves. Right:  Overploted fitted residuals, where $\overline{\sigma}$ indicates the ratio between the standard deviation of the residuals and the photon noise (see Methods for more details). The black vertical bar indicates the 1000ppm scatter level.}
	\label{fig:lightcurves}
\end{figure}

\begin{table}
    \small
    \center
    \caption{Transit depth ($(R_\mathrm{p}/R_*)^2$) for the different wavelength channels, where $R_\mathrm{p}$ is the planetary radius, $R_*$  is the stellar radius, and $\lambda_1$, $\lambda_2$ are the lower and upper edges of each wavelength channel, respectively.}
    \label{tab:spectrum}
    \begin{tabular}{c c c}
        \hline \hline
        $\lambda_1$ & $\lambda_2$ & $(R_\mathrm{p}/R_*)^2$ \\ [-2ex]
        $\mu$m & $\mu$m & ppm \\ [0.1ex]
        \hline
        $1.1390$ & $1.1730$ & $2905\pm25$ \\ [-2ex]
        $1.1730$ & $1.2040$ & $2939\pm26$ \\ [-2ex]
        $1.2040$ & $1.2330$ & $2903\pm24$ \\ [-2ex]
        $1.2330$ & $1.2610$ & $2922\pm25$ \\ [-2ex]
        $1.2610$ & $1.2890$ & $2891\pm26$ \\ [-2ex]
        $1.2890$ & $1.3160$ & $2918\pm26$ \\ [-2ex]
        $1.3160$ & $1.3420$ & $2919\pm24$ \\ [-2ex]
        $1.3420$ & $1.3690$ & $2965\pm24$ \\ [-2ex]
        $1.3690$ & $1.3960$ & $2955\pm27$ \\ [-2ex]
        $1.3960$ & $1.4230$ & $2976\pm25$ \\ [-2ex]
        $1.4230$ & $1.4500$ & $2990\pm24$ \\ [-2ex]
        $1.4500$ & $1.4780$ & $2895\pm23$ \\ [-2ex]
        $1.4780$ & $1.5060$ & $2930\pm23$ \\ [-2ex]
        $1.5060$ & $1.5340$ & $2921\pm24$ \\ [-2ex]
        $1.5340$ & $1.5620$ & $2875\pm24$ \\ [-2ex]
        $1.5620$ & $1.5910$ & $2927\pm25$ \\ [-2ex]
        $1.5910$ & $1.6200$ & $2925\pm24$ \\ 
        \hline \hline
    \end{tabular}
\end{table}

\begin{figure}
	\centering
	\includegraphics[width=0.8\textwidth]{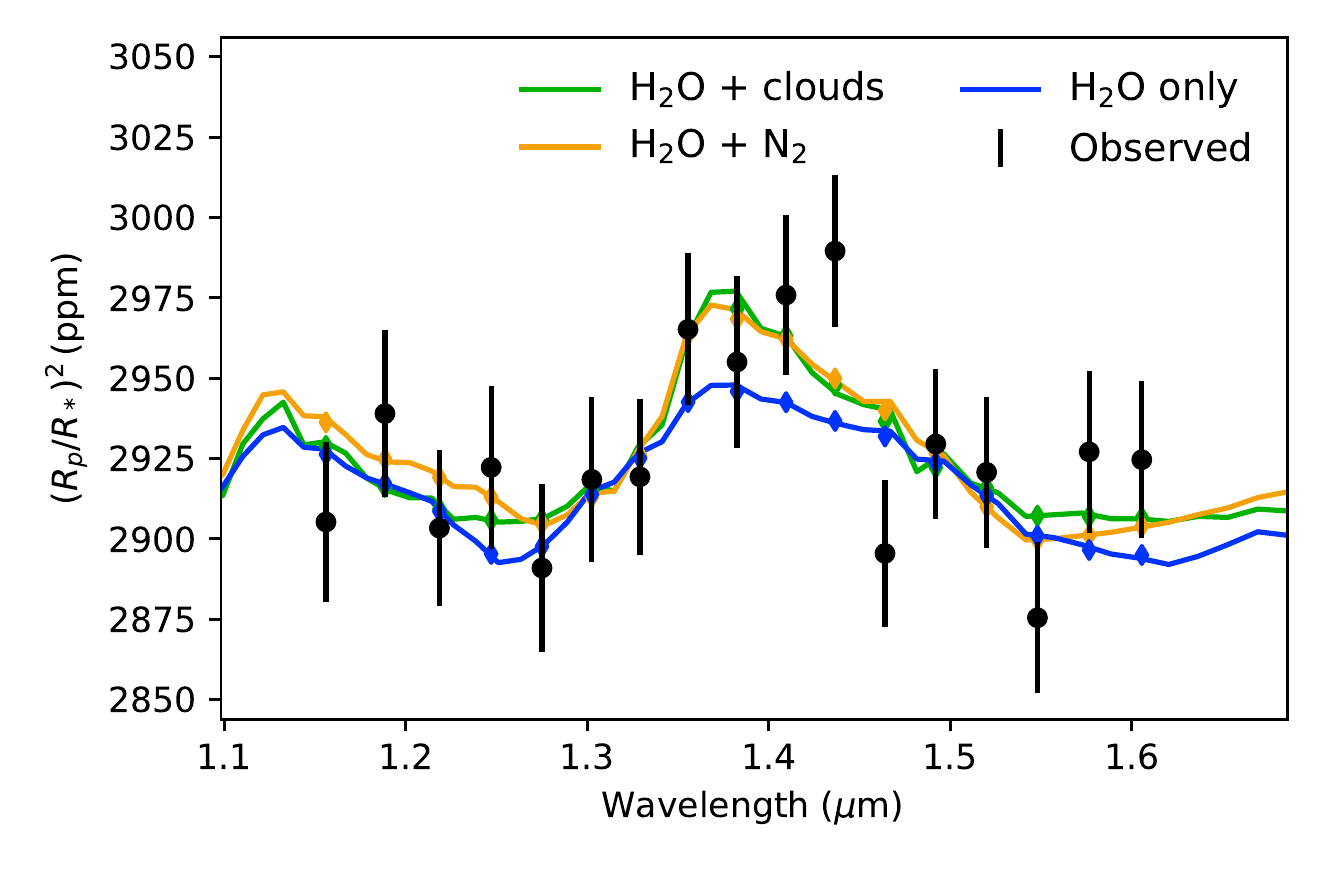}
	\includegraphics[width=0.8\textwidth]{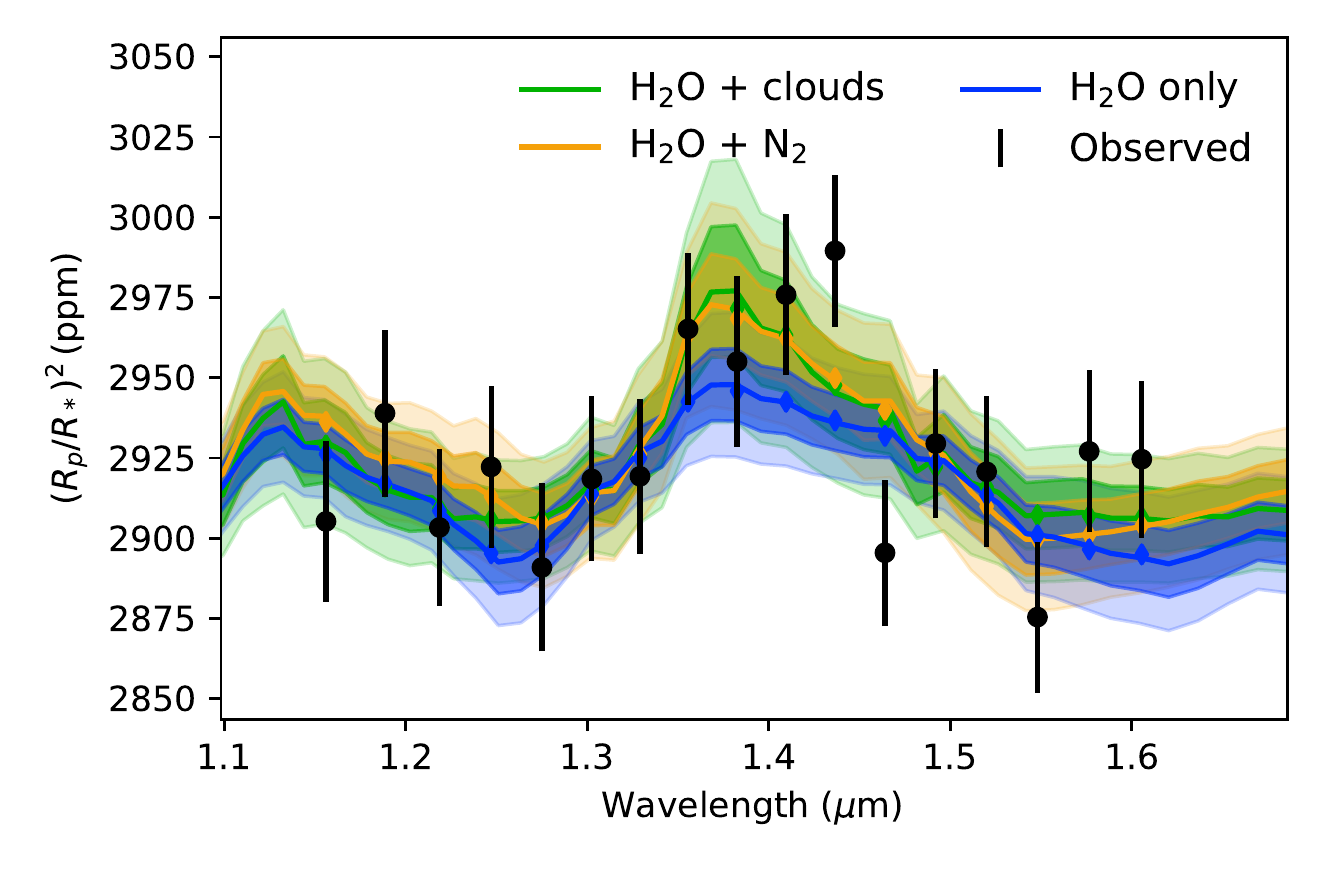}
	\caption{Best-fit models for the three different scenarios tested here: a cloud-free atmosphere containing only H$_2$O and H$_2$/He (blue), a cloud-free atmosphere containing H$_2$O, H$_2$/He and N$_2$ (orange), and a cloudy atmosphere containing only H$_2$O and H$_2$/He. (green). Top: best-fit models only. Bottom: 1$\sigma$ and 2$\sigma$ uncertainty ranges.}
	\label{fig:retrievals}
\end{figure}

\begin{figure}
	\centering
	\includegraphics[width=\textwidth]{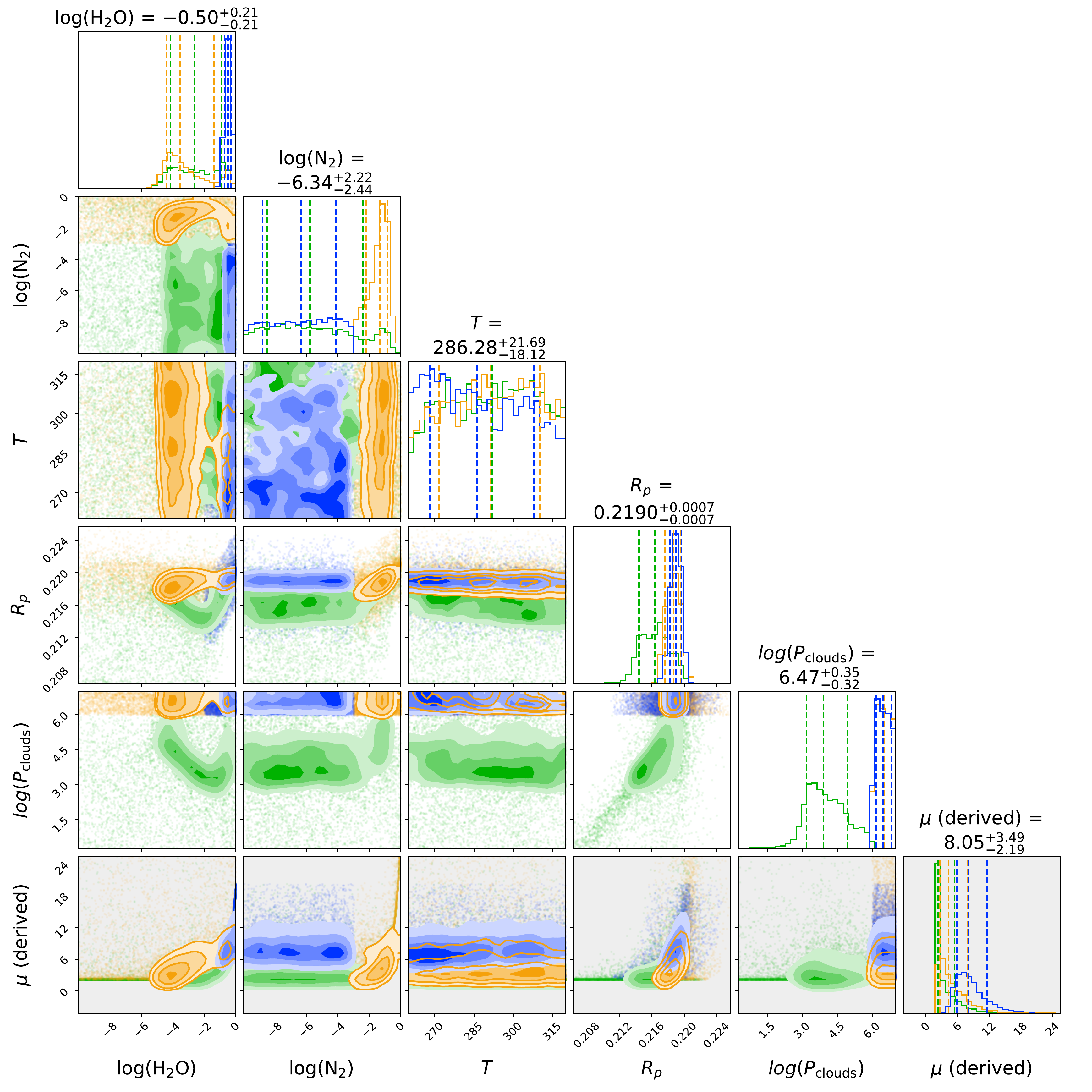}
	\caption{Posterior distributions for the three different scenarios tested here: a cloud-free atmosphere containing only H$_2$O and H$_2$/He (blue), a cloud-free atmosphere containing H$_2$O, H$_2$/He and N$_2$ (orange), and a cloudy atmosphere containing only H$_2$O and H$_2$/He. (green). The parameters shown, from top to bottom are: the volume mixing ratio of H$_2$O, the volume mixing ratio of N$_2$, the planetary temperature in K, the planetary radius in $R_\mathrm{Jup}$, the cloud top pressure in Pa, and the derived mean molecular weight.}
	\label{fig:retrievals}
\end{figure}

\newpage
\paragraph*{Methods}

\paragraph*{Observations} Nine transits of \planet\ were observed as part of the \hst\ proposals 13665 and 14682 (PI: Bj{\"o}rn Benneke) and the data are available through the MAST Archive. More specifically, the relevant HST visits are: visit 29 (06/12/2015), visit 35 (14/03/2016), and visit 30 (19/05/1016) from proposal 13665; visit 3 (02/12/2016), visit 1 (04/01/2017), visit 2 (06/02/2017), visit 4 (13/04/2017), visit 5 (30/11/2017), and visit 6 (13/05/2018) from proposal 14682. Out of these nine visits, we decided not to include the last one, as it suffered from pointing instabilities. 

Each transit was observed during five \hst\ orbits, with the \grism\ infrared grism of the \wfc\ camera (1.1 - 1.7\,$\mu$m), in the spatially scanning mode. During an exposure using the spatial scanning mode the instrument slews along the cross-dispersion direction, allowing for longer exposure times and increased signal-to-noise ratio (S/N), without the risk of saturation \cite{Deming2013}. Both forward (increasing row number) and reverse (decreasing row number) scanning were used for these observations. 

The detector settings were: SUBTYPE=SQ256SUB, SAMP\_SEQ=SPARS10, NSAMP=16,  APERTURE=GRISM256, and the scanning speed was 1.4\,$''$s$^{-1}$. The final images had a total exposure time of 103.128586 seconds, a maximum signal level of $1.9 \times 10^4$ electrons per pixel, and a total scanning length of approximately 120 pixels.  Finally, for calibration reasons, a 0.833445\,s non-dispersed (direct) image of the target was taken at the beginning of each visit, using the F130N filter and the following settings: SUBTYPE=SQ256SUB, SAMP\_SEQ=RAPID, NSAMP=4,  APERTURE=IRSUB256.

\paragraph*{Extracting the planetary spectrum} We carried out the analysis of the eight K2-18\,b transits using our specialised software for the analysis of \wfc, spatially scanned spectroscopic images\cite{Tsiaras2016B2016ApJ...832..202T, Tsiaras2016B2016ApJ...820...99T, Tsiaras2018}, which has been integrated into the \iraclis\ package (see Code Availability section). The reduction process included the following steps: zero-read subtraction, reference pixels correction, nonlinearity correction, dark current subtraction, gain conversion, sky background subtraction, calibration, flat-field correction, and bad pixels/cosmic rays correction.

We extracted the white (1.088 -- 1.68\,$\mu$m) and the spectral (Supplementary Table 1) light curves from the reduced images, taking into account the geometric distortions caused by the tilted detector of the WFC3/IR channel\cite{Tsiaras2016B2016ApJ...832..202T}. The wavelength range of the white light curve corresponds to the edges of the WFC3/G141 throughput (where the throughput drops to 30\% of the maximum). In addition, we tested two wavelength grids for the spectral light curves, with a resolving power of 20 and 50. We decided to use the latter as it was able to capture the observed water feature more precisely -- i.e there were enough data points within the wavelength range of the water feature to produce a statistically significant result.

We fitted the light curves using our transit model package \plc, the transit parameters shown in Supplementary Table 2, and limb-darkening coefficients (Supplementary Table 1) calculated based on the \phoenix\ \cite{Allard2012} model, the nonlinear formula\cite{Claret2000}, and the stellar parameters in Supplementary Table 2. 

More specifically, we fitted the white light curves with a transit model (with the planet-to-star radius ratio and the mid-transit time being the only free parameters) alongside with a model for the systematics\cite{Kreidberg2014B2014Natur.505...69K, Tsiaras2016B2016ApJ...832..202T}. It is common for WFC3 exoplanets observations to be affected by two kinds of time-dependent systematics\cite{Kreidberg2015, Evans2016, Line2016B2016AJ....152..203L, Wakeford2017B2017ApJ...835L..12W}: the long-term and short-term ``ramps''. The first affects each HST visit and has a linear behaviour, while the second affects each HST orbit and has an exponential behaviour. The formula we used for the systematics was the following:
\begin{equation}
	R_\mathrm{w}(t) = n_\mathrm{w}^\mathrm{scan}(1 - r_a (t-T_0))(1-r_{b1} e^{-r_{b2} (t-t_\mathrm{o})})
	\label{eq:white_ramp_function}
\end{equation}

\noindent where $t$ is time, $n^\mathrm{scan}_\mathrm{w}$ is a normalisation factor, $T_0$ is the mid-transit time, $t_\mathrm{o}$ is the time when each HST orbit starts, $r_a$ is the slope of a linear systematic trend along each HST visit and ($r_{b1},r_{b2}$) are the coefficients of an exponential systematic trend along each HST orbit. The normalisation factor we used was changing to $n^\mathrm{for}_\mathrm{w}$ for upwards scanning directions (forward scanning) and to $n^\mathrm{rev}_\mathrm{w}$ for downwards scanning directions (reverse scanning). The reason for using different normalisation factors is the slightly different effective exposure time due to the known up-stream/down-stream effect\cite{McCullough2012}. We, also, varied the parameters of the orbit-long exponential ramp for the first orbit in the analysed time-series ($for_{b1},for_{b2}$ instead of $r_{b1},r_{b2}$), as in many other HST observations the first orbit was affected in a different way compared to the other orbits\cite{Tsiaras2018}. While we used different ramp parameters from visit to visit, they appear to be consistent, an expected behaviour as the number of electrons collected per pixel per second is also consistent.

At a first stage we fitted the white light curves using the formulas above and the uncertainties per pixel, as propagated through the data reduction process. However, it is common in HST/WFC3 data to have additional scatter that cannot be explained by the ramp model. For this reason, we scaled-up the uncertainties on the individual data points, in order for their median to match the standard deviation of the residuals, and repeated the fitting\cite{Tsiaras2018}. From this second step of analysis, we found that the measured mid-transit times were not consistent with the expected ephemeris\cite{Benneke2017}, which we found to be: $P=32.94007\pm0.00003$\,days and $T_0=2457363.2109\pm0.0004$ BJD$_\mathrm{TDB}$\cite{Eastman2010}. Supplementary Figure 1, shows the difference between the predicted and the observed transit times using the ephemeris in the literature\cite{Benneke2017} and the one calculated in this work. We used the de-trended and time-aligned -- i.e. with TTVs removed -- white light curves to also refine the orbital parameters ($a/R_*=81.3\pm1.5$ and $i=89.56\pm0.02$\,deg). At a final step, we used all the new parameters (ephemeris, and orbital parameters) to perform a final fit on the white light curves (again having the planet-to-star radius ratio and the mid-transit time being the only free parameters).

Supplementary Figure 2 shows the raw white light curves, the detrended white light curves and the fitting residuals as well as a number of diagnostics, while Supplementary Table 3 presents the fitting results. From these, we can see that: 
\begin{itemize}
\item the final planet-to-star radius ratio is consistent among the eight different transits, demonstrating the stability of both the instrument and the analysis process,
\item on average, the white light curve residuals show an autocorrelation of 0.17, which is a low number relatively to the currently published observations of transiting exoplanets with HST\cite{Tsiaras2018} (up to 0.7), indicating a good fit,
\item uncorrected systematics are still present in the residuals which, on average, show a scatter two times larger that the expected photon noise, and
\item this extra noise component is taken into account by the increased uncertainties, as the reduced $\chi^2$ is, on average, 1.16.
\end{itemize}

Furthermore, we fitted the spectral light curves with a transit model (with the planet-to-star radius ratio being the only free parameter) alongside with a model for the systematics that included the white light curve (divide-white method\cite{Kreidberg2014B2014Natur.505...69K}), and a wavelength-dependent, visit-long slope\cite{Tsiaras2016B2016ApJ...832..202T}:
\begin{equation}
	R_\lambda(t) = n_\lambda^\mathrm{scan}(1 - \chi_\lambda (t-T_0))\frac{LC_\mathrm{w}}{M_\mathrm{w}}
	\label{eq:spectral_ramp_function}
\end{equation}

\noindent where $ \chi_\lambda$ is the slope of a wavelength-dependent linear systematic trend along each HST visit, $LC_\mathrm{w}$ is the white light curve and $M_\mathrm{w}$ is the best-fit model for the white light curve. Again, the normalisation factor we used was changed to $n^\mathrm{for}_\lambda$ for upwards scanning directions (forward scanning) and to $n^\mathrm{rev}_\lambda$ for downwards scanning directions (reverse scanning). Also, in the same way as for the white light curves, we performed an initial fit using the pipeline uncertainties and then refitted while scaling these uncertainties up, in order for their median to match the standard deviation of the residuals.

Supplementary Figures 3 to 19 show the raw spectral light curves, the detrended spectral light curves and the fitting residuals as well as a number of diagnostics, while Supplementary Table 4 presents all the fitting results and average diagnostics per spectral channel. From these, we can see that:
\begin{itemize}
\item the spectral light curves residuals show, on average, standard deviations much closer to the photon noise and lower values of autocorrelation, proving the advantage of using the white light curve as a model compared to the ramp model, and
\item any extra noise component is taken into account by the scaled-up uncertainties, as the reduced $\chi^2$ is for all channels, on average, close to unity.
\end{itemize}

Finally, the eight spectra of \planet\ (Supplementary Figure 20) were combined, using a weighted average, to produce the final spectrum (Table \ref{tab:spectrum}).

\paragraph*{Stellar contamination} K2-18 is a moderately active M2.5V star, with a variability of 1.7\% in the
B band and 1.38\% in the R band\cite{Sarkis2018}. Hence, to make sure that the observed water feature is not the effect of stellar contamination we fitted the observed spectrum with a model that assumes a flat planetary spectrum and contribution only from the star (M2V star as described in Rackham et al. 2018\cite{Rackham2018}). The model that best describes our data has a spot coverage of 26\% and a faculae coverage of 73\%. We plot this spectrum versus the observed one in Supplementary Figure 21. In addition, we plot the spectrum produced by the spot and faculae combination reported by Rackham et al. 2018\cite{Rackham2018} and correspond to a 1\% I-band variability, for reference. However, as Supplementary Figure 21 shows, the best-fit model cannot describe the observed water feature. From these we conclude that there is no combination of stellar properties that could introduce the observed water feature.

\paragraph*{Atmospheric retrieval} We fitted the final planetary spectrum using our Bayesian atmospheric retrieval framework \taurex \cite{Waldmann2015B2015ApJ...813...13W, Waldmann2015B2015ApJ...802..107W}, which fully maps the correlations between the fitted atmospheric parameters through nested sampling\cite{Skilling2006, Feroz2009}. 

The atmosphere of \planet\ was simulated as a plane-parallel atmosphere with pressures ranging from 10$^{-4}$ to 10$^6$\,Pa, sampled uniformly in log-space by 100 atmospheric layers, assuming an isothermal temperature-pressure profile. We initially tested fitting for a number of trace-gases -- H$_2$O\cite{Barber2006}, CO\cite{Rothman2010}, CO$_2$\cite{Rothman2010}, CH$_4$\cite{Yurchenko2014} and NH$_3$\cite{Yurchenko2011} -- but found only water vapour to play a significant role. Hence, we proceeded only with this molecule. We also included the effect of clouds using a grey/flat-line model, as the quality and wavelength ranges of the currently available observations do not allow us to make any reasonable constraints on the haze properties of the planet. Finally, we included the spectroscopically inactive N$_2$ as an inactive gas, to account for any unseen absorbers -- e.g. methane, which is expected at these temperatures. As free parameters in our models we had: the volume mixing ratio of H$_2$O (log-uniform prior between 10$^{-10}$ and 1.0), the volume mixing ratio of N$_2$ (log-uniform prior between 10$^{-10}$ and 1.0), the planetary temperature (uniform prior between 260 and 320 K), the planetary radius (uniform prior between 0.05 and 0.5 $R_\mathrm{Jup}$, and the cloud top pressure (log-uniform prior between 10$^{-3}$ and 10$^{7}$ Pa, where 10$^{7}$ Pa represents a cloud free atmosphere). We restricted the temperature prior compared to all the possible temperatures for different values of albedo and emissivity because, since we can detect only water, the temperature of atmospheric part probed must be higher than the freezing point of water ($\sim$260\,K at 1\,mbar).

We identified three solutions: a) a cloud-free atmosphere containing only H$_2$O and H$_2$/He, b) a cloud-free atmosphere containing H$_2$O, H$_2$/He and N$_2$, and c) a cloudy atmosphere containing only H$_2$O and H$_2$/He. The best-fit spectra and the posterior plots are shown in Figure \ref{fig:retrievals}. In all cases, a statistically significant atmosphere around \planet\ was retrieved with an ADI\cite{Tsiaras2018} of 5.0, 4.7 and 4.0, respectively. The ADI is the positively defined logarithmic Bayes Factor, where the null hypothesis is a model that contains no active trace gases, Rayleigh scattering or collision induced absorption -- i.e. a flat spectrum. An ADI of 5 corresponds to approximately a 3.6$\sigma$\cite{Benneke2013, Waldmann2015B2015ApJ...802..107W} detection of an atmosphere. The values are too similar to distinguish between the three scenarios.

\paragraph*{Data Availability} The data analysed in this work are available 
through the NASA MAST HST archive 
(\href{https://archive.stsci.edu/}{https://archive.stsci.edu}) programs 13665 and 14682. The molecular line lists used are available from the ExoMol webpage (\href{www.exomol.com}{www.exomol.com}). The final 
and intermediate results (reduced data, extracted light curves, light curve fitting results and atmospheric fitting results) are available through the UCL-Exoplanets webpage 
(\href{https://www.ucl.ac.uk/exoplanets}{https://www.ucl.ac.uk/exoplanets}).

\paragraph*{Code Availability} All the software used to produced the presented 
results are publicly available through the UCL-Exoplanets GitHub page 
(\href{https://github.com/ucl-exoplanets/}{https://github.com/ucl-exoplanets}). 
More specifically, the codes used were:
\vspace{-1.0cm}
\begin{itemize}
\item \taurex\ 
(\href{https://github.com/ucl-exoplanets/TauREx_public}{https://github.com/ucl-exoplanets/TauREx\_public}), \vspace{-0.4cm}
\item \iraclis\ 
(\href{https://github.com/ucl-exoplanets/Iraclis}{https://github.com/ucl-exoplanets/Iraclis}), and \vspace{-0.4cm}
\item \plc\ 
(\href{https://github.com/ucl-exoplanets/pylightcurve}{https://github.com/ucl-exoplanets/pylightcurve}).
\end{itemize}



\newpage
\nolinenumbers
\pagenumbering{gobble}
\begin{center}
\textbf{\Large{Water vapour in the atmosphere of the habitable-zone eight Earth-mass planet K2-18 b}}
\end{center}
\newpage
\setcounter{table}{0}
\setcounter{figure}{0}
\makeatletter
\renewcommand{\fnum@figure}{Supplementary~\figurename~\thefigure}
\renewcommand{\fnum@table}{Supplementary~\tablename~\thetable}
\makeatother

\newpage

\begin{table}
    \small
    \center
    \caption{Limb-darkening coefficients for the different wavelength channels.}
    \label{tab:ldcoeff}
    \begin{tabular}{c c c c c c}
        \hline \hline
        $\lambda_1$ & $\lambda_2$ & $a_1$ & $a_2$ & $a_3$ & $a_4$ \\ [-2ex]
        $\mu$m & $\mu$m \\ [0.1ex]
        \hline
        $1.1390$ & $1.1730$ & $1.7770$ & $-2.3809$ & $1.9836$ & $-0.6622$ \\ [-2ex]
        $1.1730$ & $1.2040$ & $1.7751$ & $-2.3766$ & $1.9665$ & $-0.6526$ \\ [-2ex]
        $1.2040$ & $1.2330$ & $1.8060$ & $-2.5127$ & $2.1320$ & $-0.7193$ \\ [-2ex]
        $1.2330$ & $1.2610$ & $1.8025$ & $-2.5249$ & $2.1512$ & $-0.7279$ \\ [-2ex]
        $1.2610$ & $1.2890$ & $1.8355$ & $-2.6551$ & $2.3012$ & $-0.7866$ \\ [-2ex]
        $1.2890$ & $1.3160$ & $1.8596$ & $-2.7798$ & $2.4530$ & $-0.8480$ \\ [-2ex]
        $1.3160$ & $1.3420$ & $1.7152$ & $-2.3020$ & $1.9119$ & $-0.6377$ \\ [-2ex]
        $1.3420$ & $1.3690$ & $1.4322$ & $-1.3839$ & $0.8815$ & $-0.2404$ \\ [-2ex]
        $1.3690$ & $1.3960$ & $1.3984$ & $-1.2141$ & $0.6422$ & $-0.1360$ \\ [-2ex]
        $1.3960$ & $1.4230$ & $1.2459$ & $-0.8054$ & $0.2350$ & $0.0087$ \\ [-2ex]
        $1.4230$ & $1.4500$ & $1.1758$ & $-0.5644$ & $-0.0498$ & $0.1203$ \\ [-2ex]
        $1.4500$ & $1.4780$ & $1.1154$ & $-0.4431$ & $-0.1382$ & $0.1431$ \\ [-2ex]
        $1.4780$ & $1.5060$ & $1.1652$ & $-0.5113$ & $-0.1367$ & $0.1590$ \\ [-2ex]
        $1.5060$ & $1.5340$ & $1.2015$ & $-0.5856$ & $-0.0719$ & $0.1388$ \\ [-2ex]
        $1.5340$ & $1.5620$ & $1.2075$ & $-0.5973$ & $-0.0613$ & $0.1353$ \\ [-2ex]
        $1.5620$ & $1.5910$ & $1.3457$ & $-0.9365$ & $0.2415$ & $0.0362$ \\ [-2ex]
        $1.5910$ & $1.6200$ & $1.4800$ & $-1.2998$ & $0.6065$ & $-0.0941$ \\ [-2ex]
        $1.0880$ & $1.6800$ & $1.5202$ & $-1.5934$ & $1.0698$ & $-0.3030$ \\ 
        \hline \hline
    \end{tabular}
\end{table}

\newpage

\begin{table}
	\small
	\center
	\caption{Observationally determined parameters of \planet.}
	\label{tab:parameters}
	\begin{tabular}{c | c }
		
		\hline \hline
		\multicolumn{2}{c}{Stellar parameters } 				
							\\ [0.1ex]
		\hline
		$\mathrm{[Fe/H]} \, \mathrm{[dex]}$\cite{Benneke2017}	& 0.123 
$\pm$ 0.157						\\ [-2ex]
		$T_\mathrm{eff} \, \mathrm{[K]}$\cite{Benneke2017}		
& 3457 $\pm$ 39						\\ [-2ex]
		$M_* \, [M_{\odot}]$\cite{Benneke2017}				
& 0.359 $\pm$ 0.047						\\ [-2ex]
		$R_* \, [R_{\odot}]$\cite{Benneke2017}				
& 0.411 $\pm$ 0.038						\\ [1.0ex]
		
		\hline \hline
		\multicolumn{2}{c}{Planetary parameters}				
							\\ [0.1ex]
		\hline
		$T_\mathrm{eq} \, \mathrm{[K]}$\cite{Cloutier2017} 		
& 235 $\pm$ 9	\\ [-2ex]
		Bond albedo = 0.3		&  						\\ [-1ex]
		$M_\mathrm{p} \, [M_\oplus]$\cite{Cloutier2017}		& 7.96 
$\pm$ 1.91						\\ [-1ex]
		$R_\mathrm{p} \, [R_\oplus]$\cite{Benneke2017}		& 
2.279$_{-0.025}^{+0.026}$				\\ [-1ex]
		$a \, \mathrm{[AU]}$\cite{Benneke2017} 				
& 0.1429$_{-0.0065}^{+0.0060}$			\\ [1.0ex]
		
		\hline\hline
		\multicolumn{2}{c}{Transit parameters - this study}					
						\\ [0.1ex]
		\hline
		$T_0 \, \mathrm{[days, BJD_{TDB}]}$			& 2457363.2109 $\pm$ 0.0004	\\ [-1ex]
		$\mathrm{Period} \,  \mathrm{[days]}$ 			& 32.94007 $\pm$ 0.00003	\\ [-1ex]
		$R\mathrm{p}/R_*\, \mathrm{(WFC3-G141 \ band)}$ & 0.05405 $\pm$ 0.00014		\\ [-1ex]
		$a/R_*$									& 81.3 $\pm$ 1.5			\\ [-1ex]
		$i \, \mathrm{[deg]}$							& 89.56 $\pm$ 0.02			\\					
				
\hline\hline								
	\end{tabular}
\end{table}

\newpage

\begin{table}
    \small
    \center
    \caption{Fitting results and residuals diagnostics for the eight different white light curves ($R^2$: autocorelation, $\overline{\sigma}$: standard deviation relative to photon noise, $\overline{\chi}^2$: reduced ${\chi}^2$ calculated with the rescaled uncertainties).}
    \label{tab:white_results}
    \begin{tabular}{c c c c c c}
        \hline \hline
        $T_0$ & $T_0$ & $(R_p/R_*)^2$ & $R^2$ & $\overline{\sigma}$ & $\overline{\chi}^2$ \\ [-2ex]
        HJD$_\mathrm{UTC}$ & BJD$_\mathrm{TDB}$ & ppm \\ [0.1ex]
        \hline
        $2457363.2101_{-0.0002}^{+0.0003}$ & $2457363.210869_{-0.0002}^{+0.0003}$ &$2928_{-39}^{+39}$ & $0.07$ & $1.84$ & $1.14$ \\
        $2457462.02922_{-0.00017}^{+0.00023}$ & $2457462.029990_{-0.00017}^{+0.00023}$ & $2873_{-33}^{+38}$ & $0.11$ & $2.1$ & $1.14$ \\
        $2457527.9107_{-0.0002}^{+0.0002}$ & $2457527.911470_{-0.0002}^{+0.0002}$ & $2944_{-43}^{+49}$ & $0.17$ & $2.04$ & $1.14$ \\
        $2457725.5504_{-0.0002}^{+0.0003}$ & $2457725.551173_{-0.0002}^{+0.0003}$ & $2931_{-25}^{+33}$ & $0.21$ & $1.78$ & $1.16$ \\
        $2457758.49096_{-0.00016}^{+0.00021}$ & $2457758.49096_{-0.00016}^{+0.00021}$ & $2968_{-45}^{+51}$ & $0.27$ & $2.1$ & $1.15$ \\
        $2457791.4315_{-0.0004}^{+0.0004}$ & $2457791.432286_{-0.0004}^{+0.0004}$ & $2921_{-47}^{+54}$ & $0.24$ & $2.39$ & $1.19$ \\
        $2457857.31143_{-0.00016}^{+0.00016}$ & $2457857.312218_{-0.00016}^{+0.00016}$ & $2918_{-36}^{+41}$ & $0.15$ & $1.84$ & $1.13$ \\
        $2458087.8912_{-0.0003}^{+0.0003}$ & $2458087.891996_{-0.0003}^{+0.0003}$ & $2888_{-55}^{+83}$ & $0.16$ & $2.55$ & $1.22$ \\
        \hline \hline
    \end{tabular}
\end{table}

\newpage

\begin{table}
    \small
    \center
    \caption{Fitting results and residuals diagnostics for the different wavelength channels ($R^2$: autocorrelation, $\overline{\sigma}$: standard deviation relative to photon noise, $\overline{\chi}^2$: reduced ${\chi}^2$). The diagnostics are presented as the mean and the standard deviation of the eight visits. Information on individual visits can be found in Supplementary Figures 3 to 19. }
    \label{tab:spectral_results}
    \begin{tabular}{c c c c c c}
        \hline \hline
        $\lambda_1$ & $\lambda_2$ & $(R_\mathrm{p}/R_*)^2$ & $R^2$ & $\overline{\sigma}$ & $\overline{\chi}^2$ \\ [-2ex]
        $\mu$m & $\mu$m & ppm & $\mu \, (\sigma)$ & $\mu \, (\sigma)$ & $\mu \, (\sigma)$ \\ [0.1ex]
        \hline
        $1.1390$ & $1.1730$ & $2905\pm25$ & $0.13\,(0.08)$ & $1.09\,(0.08)$ & $1.060\,(0.007)$ \\ [-2ex]
        $1.1730$ & $1.2040$ & $2939\pm26$ & $0.13\,(0.04)$ & $1.12\,(0.10)$ & $1.060\,(0.006)$ \\ [-2ex]
        $1.2040$ & $1.2330$ & $2903\pm24$ & $0.13\,(0.05)$ & $1.07\,(0.05)$ & $1.061\,(0.007)$ \\ [-2ex]
        $1.2330$ & $1.2610$ & $2922\pm25$ & $0.15\,(0.06)$ & $1.08\,(0.10)$ & $1.061\,(0.007)$ \\ [-2ex]
        $1.2610$ & $1.2890$ & $2891\pm26$ & $0.14\,(0.08)$ & $1.16\,(0.17)$ & $1.059\,(0.008)$ \\ [-2ex]
        $1.2890$ & $1.3160$ & $2918\pm26$ & $0.11\,(0.06)$ & $1.13\,(0.16)$ & $1.061\,(0.006)$ \\ [-2ex]
        $1.3160$ & $1.3420$ & $2919\pm24$ & $0.15\,(0.07)$ & $1.09\,(0.13)$ & $1.061\,(0.007)$ \\ [-2ex]
        $1.3420$ & $1.3690$ & $2965\pm24$ & $0.13\,(0.05)$ & $1.05\,(0.08)$ & $1.061\,(0.007)$ \\ [-2ex]
        $1.3690$ & $1.3960$ & $2955\pm27$ & $0.15\,(0.09)$ & $1.17\,(0.07)$ & $1.062\,(0.007)$ \\ [-2ex]
        $1.3960$ & $1.4230$ & $2976\pm25$ & $0.16\,(0.06)$ & $1.10\,(0.08)$ & $1.061\,(0.006)$ \\ [-2ex]
        $1.4230$ & $1.4500$ & $2990\pm24$ & $0.12\,(0.09)$ & $1.01\,(0.08)$ & $1.061\,(0.007)$ \\ [-2ex]
        $1.4500$ & $1.4780$ & $2895\pm23$ & $0.18\,(0.06)$ & $1.02\,(0.09)$ & $1.061\,(0.006)$ \\ [-2ex]
        $1.4780$ & $1.5060$ & $2930\pm23$ & $0.15\,(0.07)$ & $1.04\,(0.08)$ & $1.061\,(0.007)$ \\ [-2ex]
        $1.5060$ & $1.5340$ & $2921\pm24$ & $0.12\,(0.06)$ & $1.05\,(0.08)$ & $1.062\,(0.006)$ \\ [-2ex]
        $1.5340$ & $1.5620$ & $2875\pm24$ & $0.15\,(0.07)$ & $1.02\,(0.06)$ & $1.061\,(0.006)$ \\ [-2ex]
        $1.5620$ & $1.5910$ & $2927\pm25$ & $0.13\,(0.09)$ & $1.10\,(0.08)$ & $1.061\,(0.006)$ \\ [-2ex]
        $1.5910$ & $1.6200$ & $2925\pm24$ & $0.13\,(0.06)$ & $1.14\,(0.16)$ & $1.063\,(0.009)$ \\ 
        \hline \hline
    \end{tabular}
\end{table}

\newpage

\begin{figure}[h!]
	\centering
	\includegraphics[width=\textwidth]{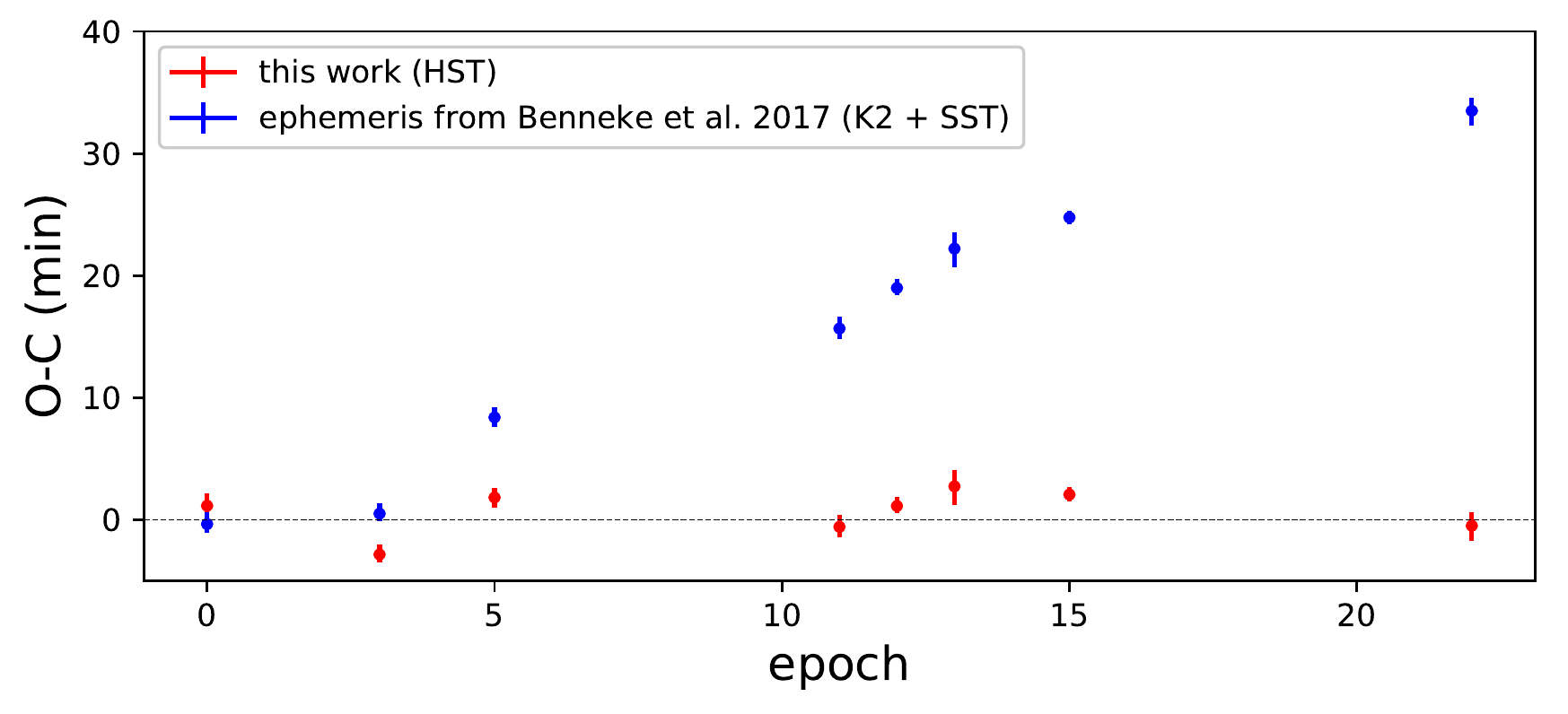}
	\caption{O-C diagram using the ephemeris from Benneke et al. 2017 in blue ($P=32.939622_{0.000094}^{0.000099}$\,days and $T_0=2457264.39135_{0.00066}^{0.00062}$ BJD$_\mathrm{UTC}$) and our updated ephemeris based on the \hst\ data in red ($P=32.94007\pm0.00003$\,days and $T_0=2457363.2109\pm0.0004$ BJD$_\mathrm{TDB}$).}
	\label{fig:oc}
\end{figure}

\newpage

\begin{figure}
	\centering
	\includegraphics[width=\textwidth]{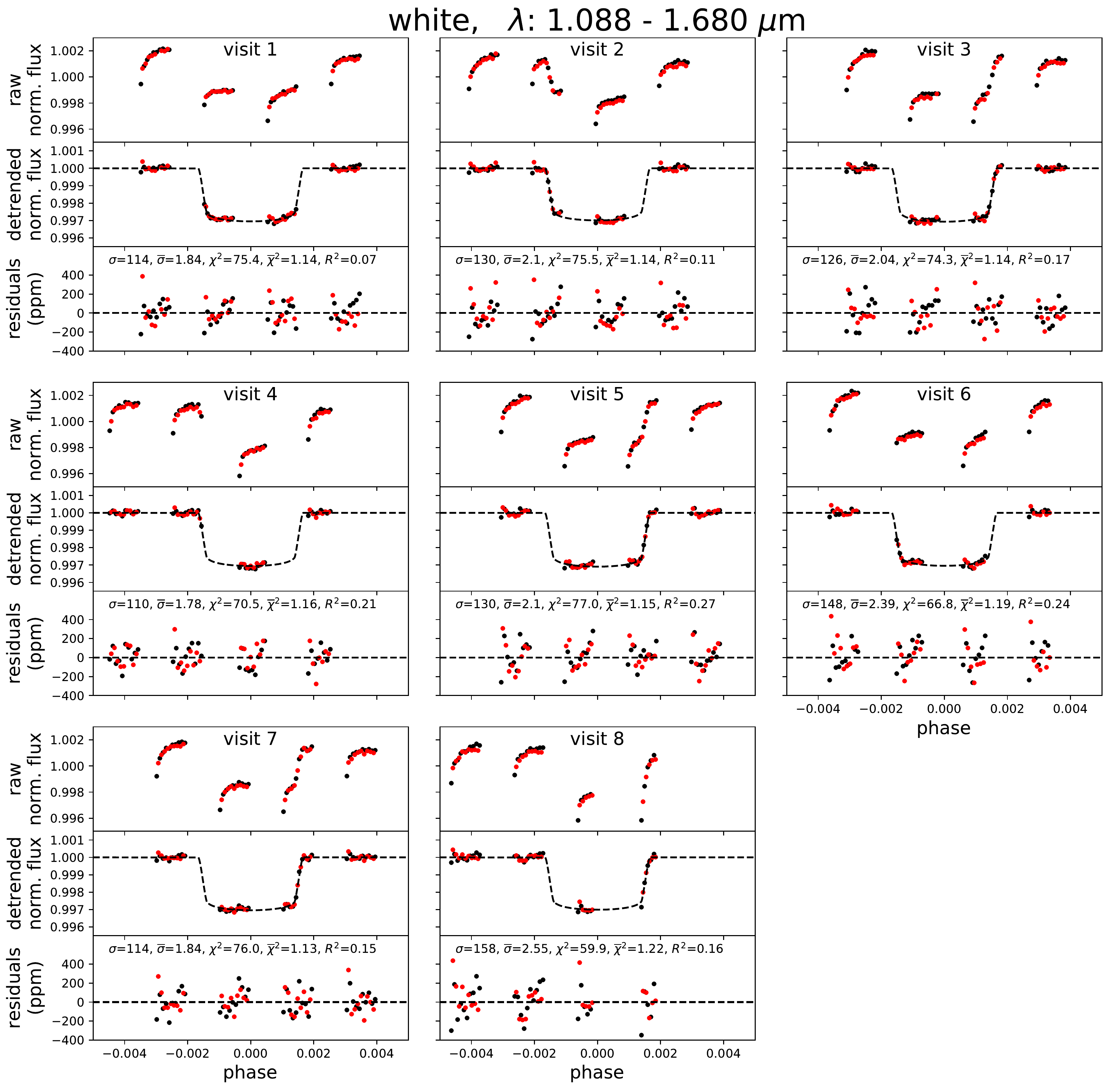}
	\caption{Analysis of the eight \planet\ white light curves, for the forward (black) and reverse (red) scans. In each subplot, from top to bottom: A) Raw light curves extracted from the HST observations. B) Raw light curves divided by the best-fit model for the systematics. C) Fitting residuals where their standard deviation ($\sigma$), their standard deviation relatively to the photon noise ($\overline{\sigma}$), their chi squared ($\chi^2$), their reduced chi squared ($\overline{\chi}^2$), and their autocorrelation ($R^2$).}
	\label{fig:white_fit}
\end{figure}

\newpage
\begin{figure}
\centering
\includegraphics[width=\textwidth]{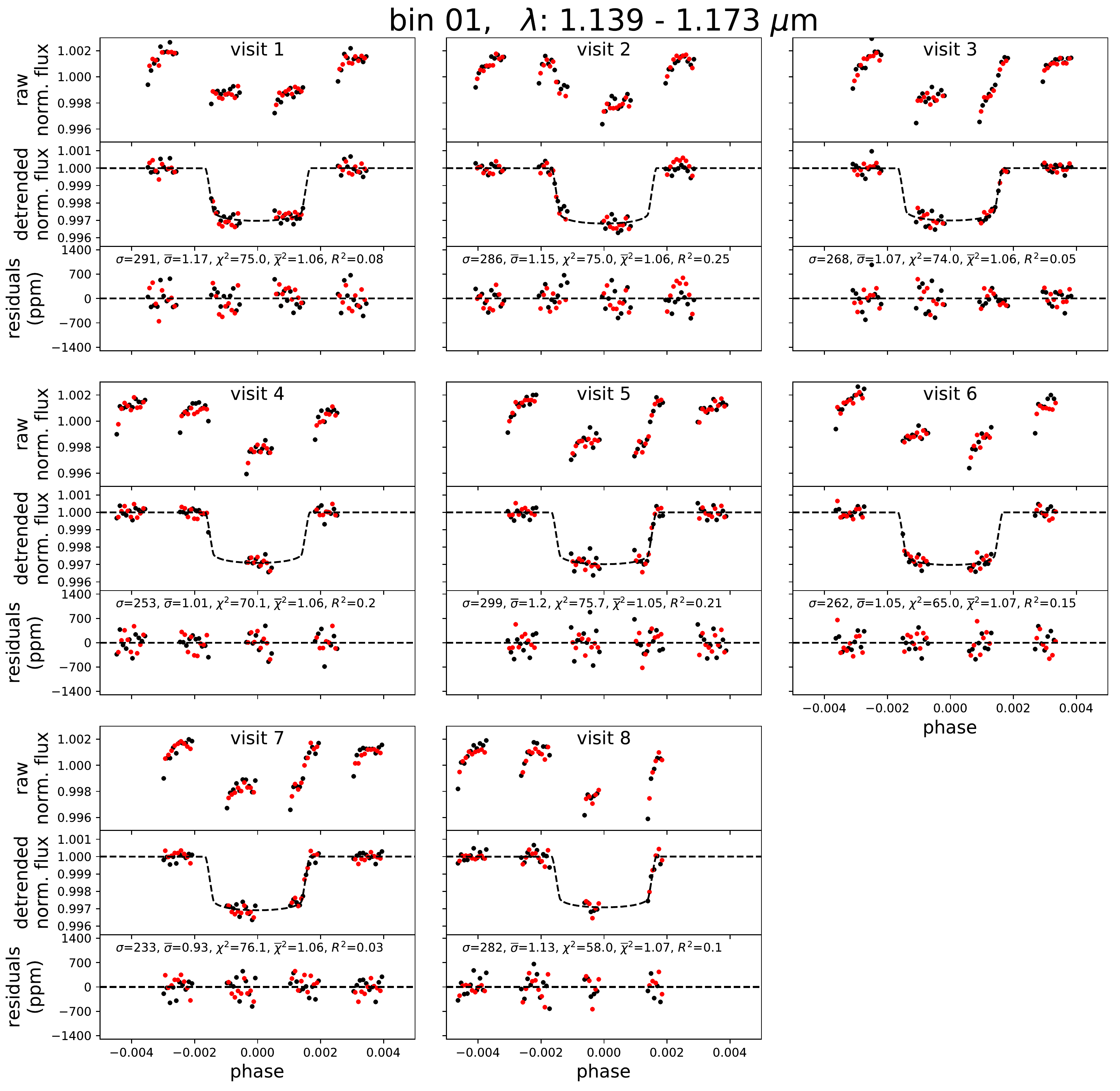}
\caption{Same as Supplementary Figure 2 for the spectral channel between 1.139 and 1.173 $\mu$m. }
\label{fig:bin_01_fit}
\end{figure}

\newpage
\begin{figure}
\centering
\includegraphics[width=\textwidth]{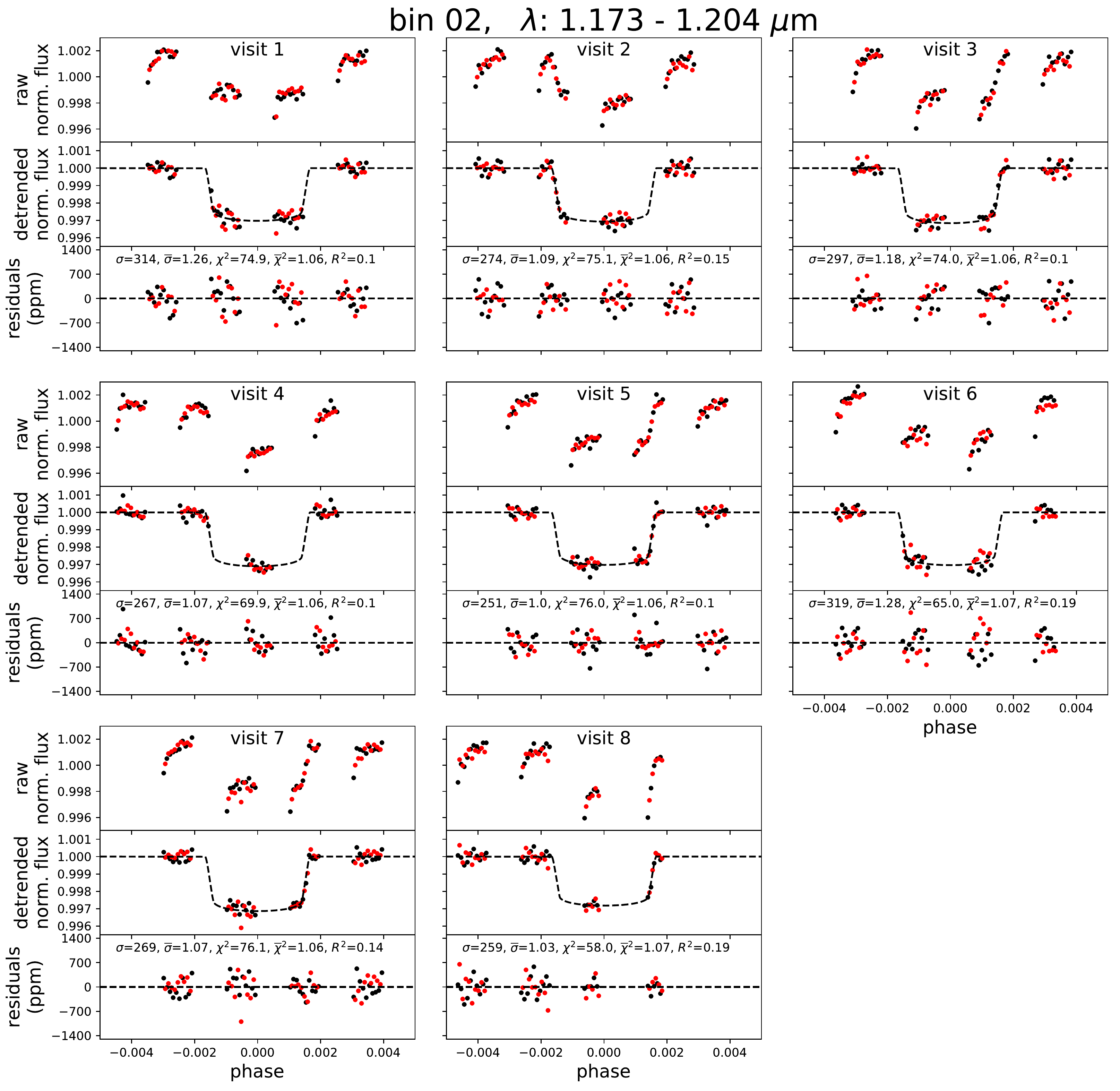}
\caption{Same as Supplementary Figure 2 for the spectral channel between 1.173 and 1.204 $\mu$m. }
\label{fig:bin_02_fit}
\end{figure}

\newpage
\begin{figure}
\centering
\includegraphics[width=\textwidth]{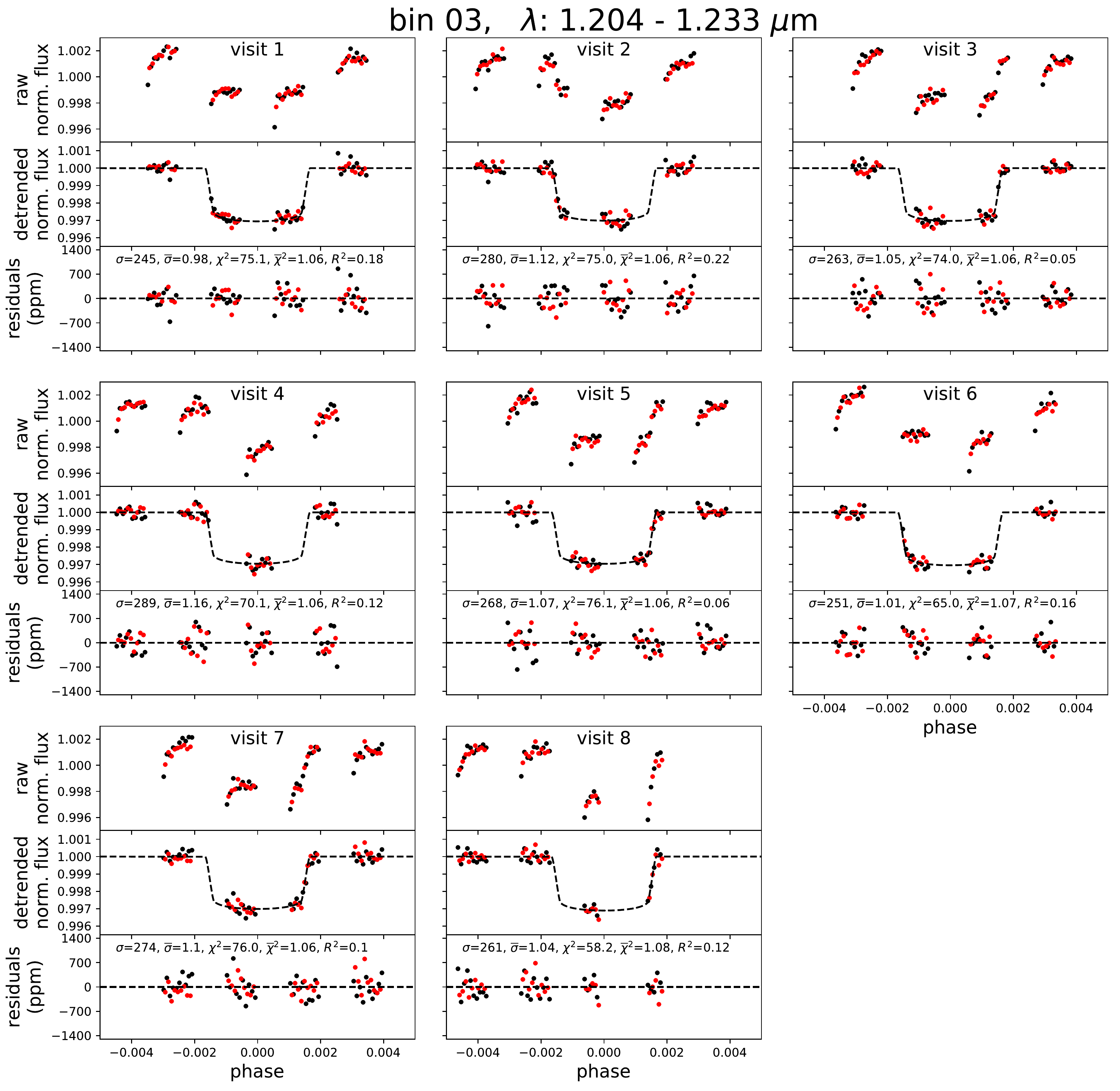}
\caption{Same as Supplementary Figure 2 for the spectral channel between 1.204 and 1.233 $\mu$m. }
\label{fig:bin_03_fit}
\end{figure}

\newpage
\begin{figure}
\centering
\includegraphics[width=\textwidth]{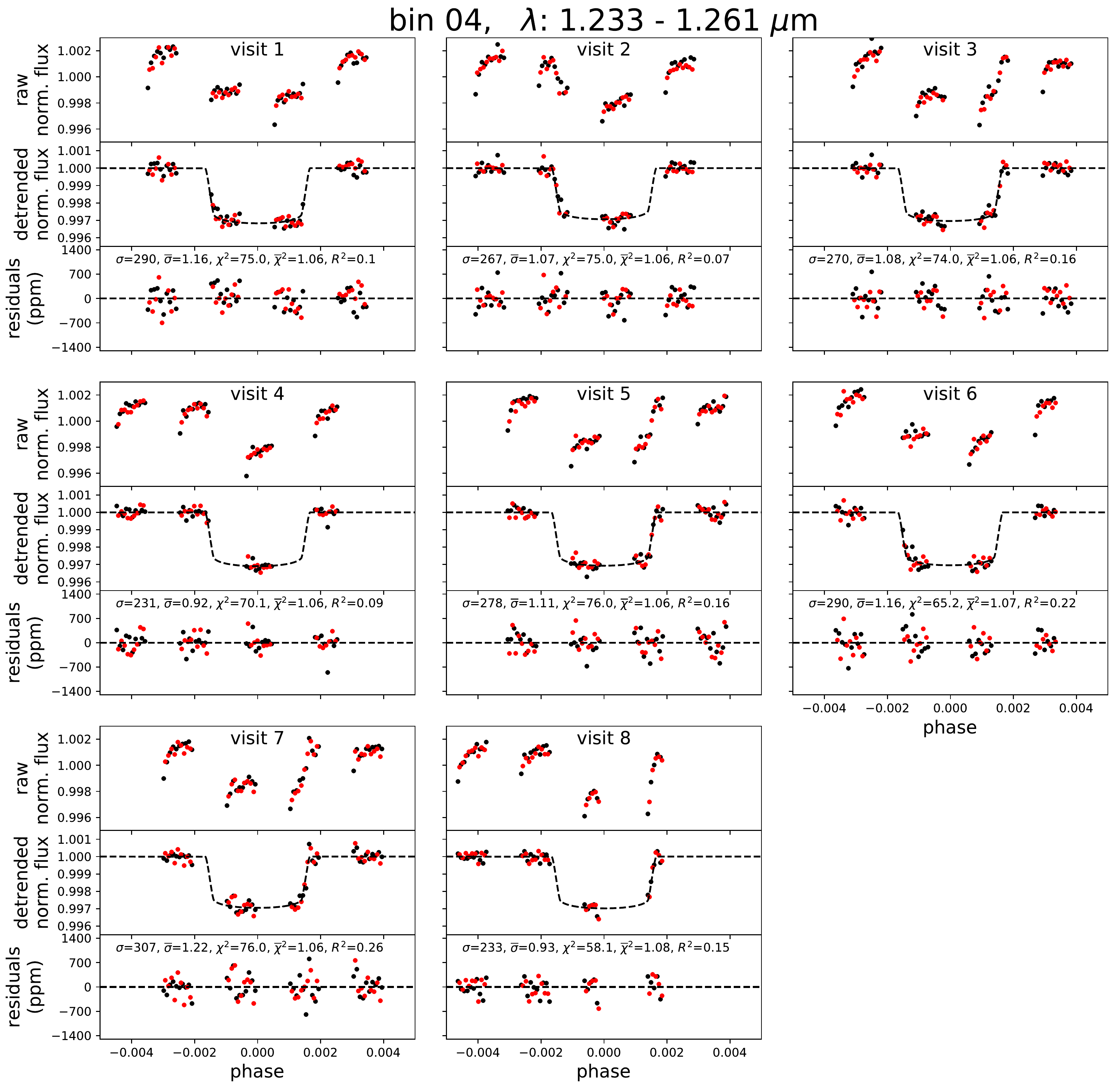}
\caption{Same as Supplementary Figure 2 for the spectral channel between 1.233 and 1.261 $\mu$m. }
\label{fig:bin_04_fit}
\end{figure}

\newpage
\begin{figure}
\centering
\includegraphics[width=\textwidth]{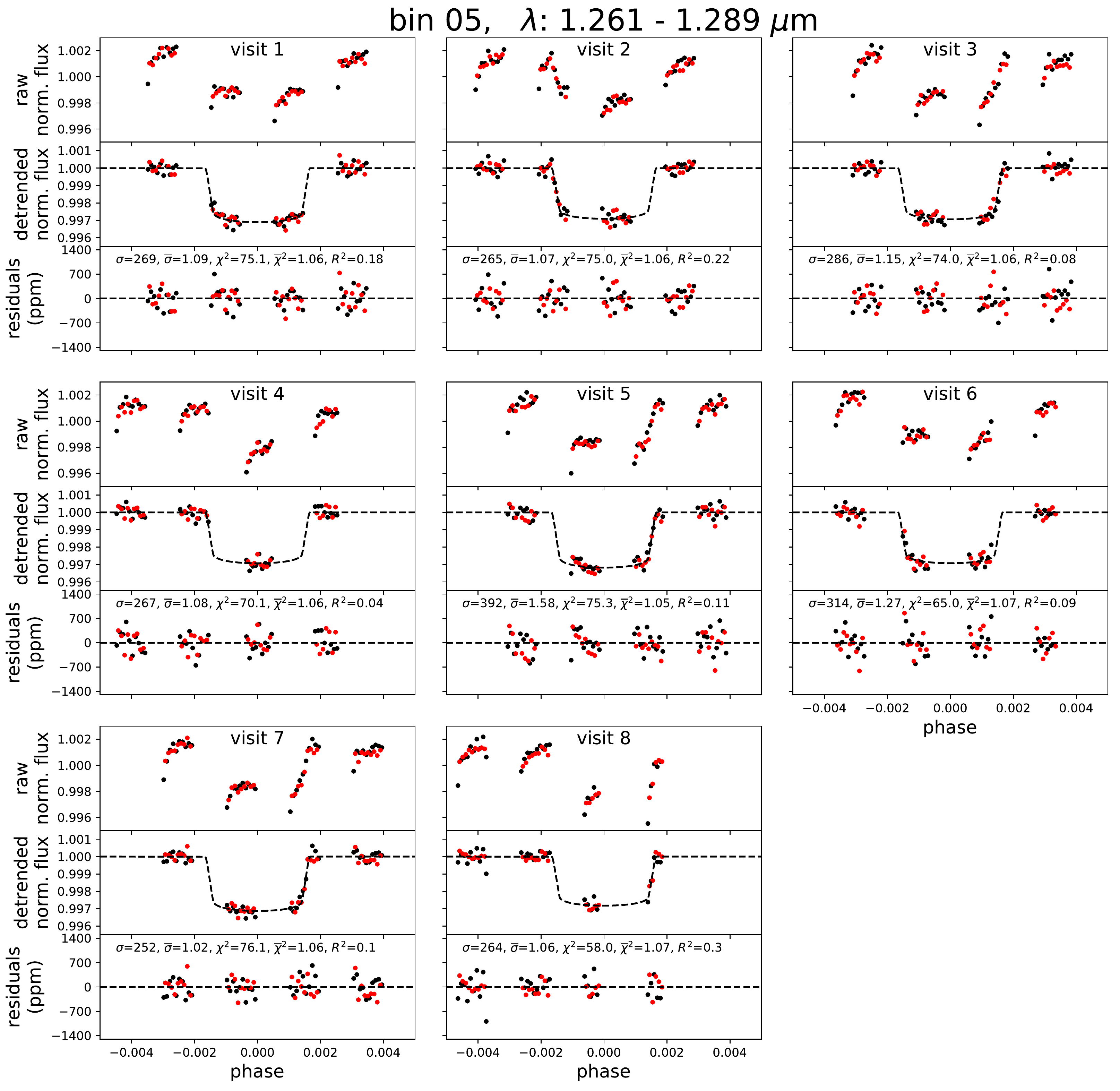}
\caption{Same as Supplementary Figure 2 for the spectral channel between 1.261 and 1.289 $\mu$m. }
\label{fig:bin_05_fit}
\end{figure}

\newpage
\begin{figure}
\centering
\includegraphics[width=\textwidth]{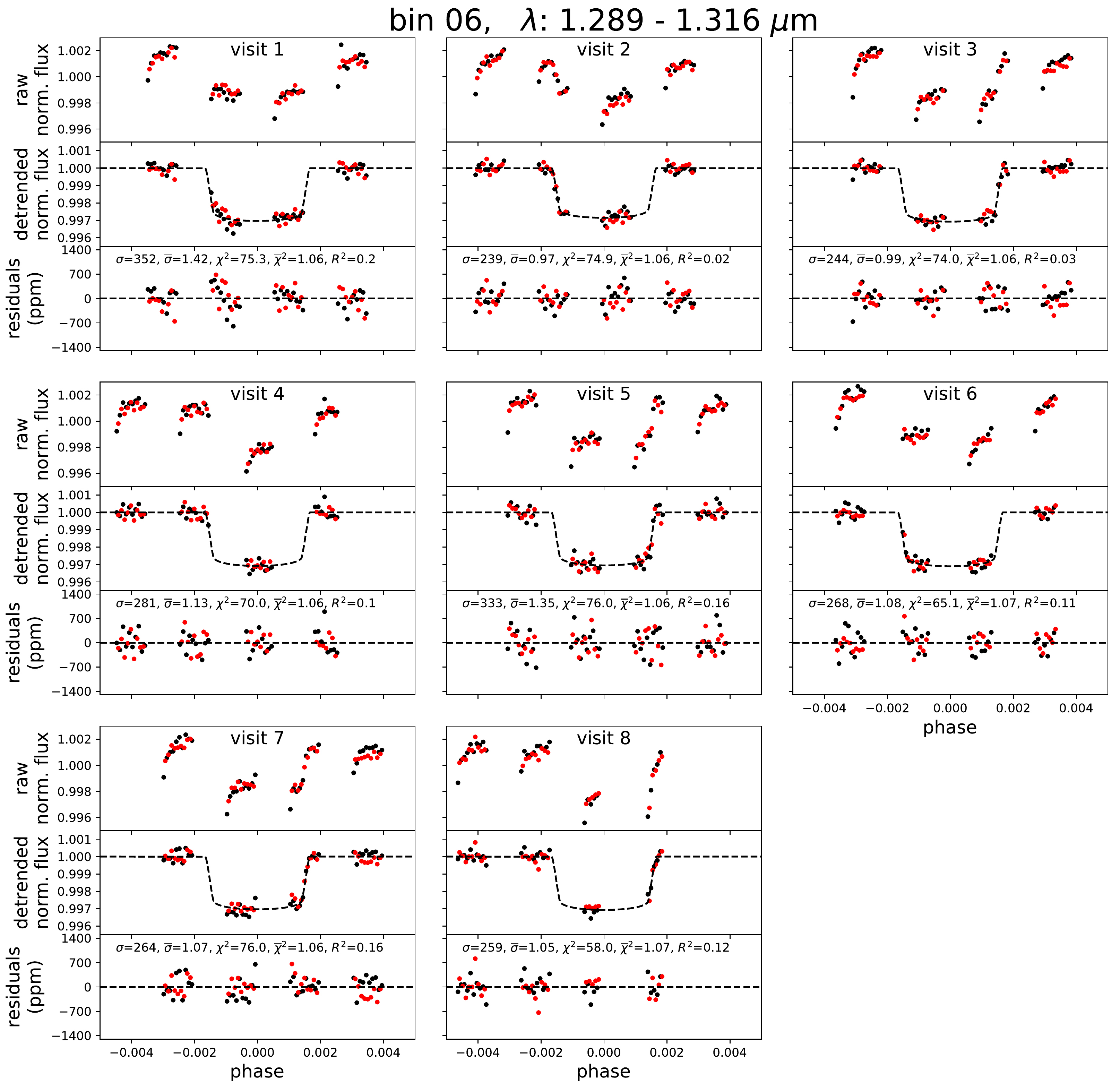}
\caption{Same as Supplementary Figure 2 for the spectral channel between 1.289 and 1.316 $\mu$m. }
\label{fig:bin_06_fit}
\end{figure}

\newpage
\begin{figure}
\centering
\includegraphics[width=\textwidth]{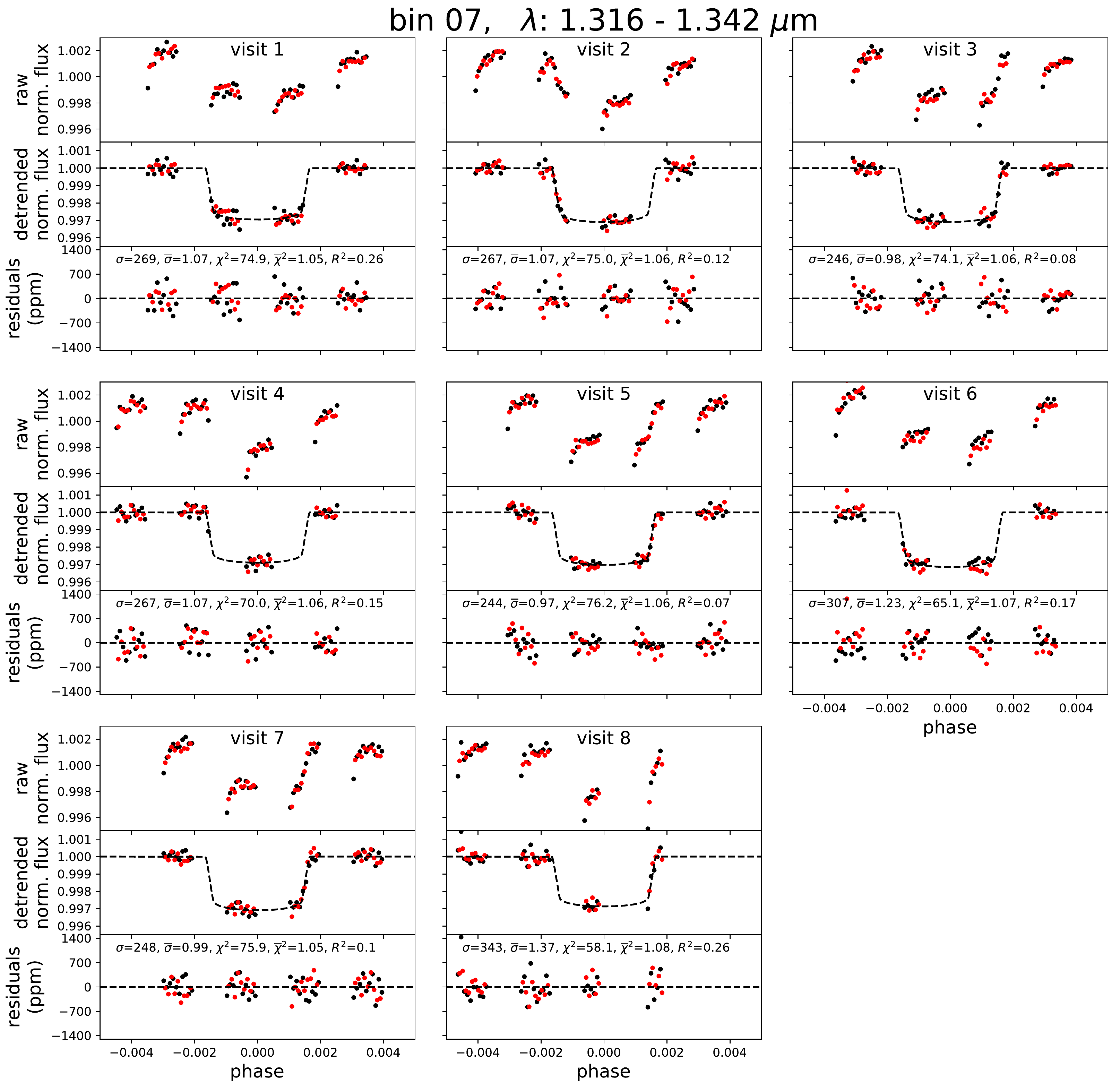}
\caption{Same as Supplementary Figure 2 for the spectral channel between 1.316 and 1.342 $\mu$m. }
\label{fig:bin_07_fit}
\end{figure}

\newpage
\begin{figure}
\centering
\includegraphics[width=\textwidth]{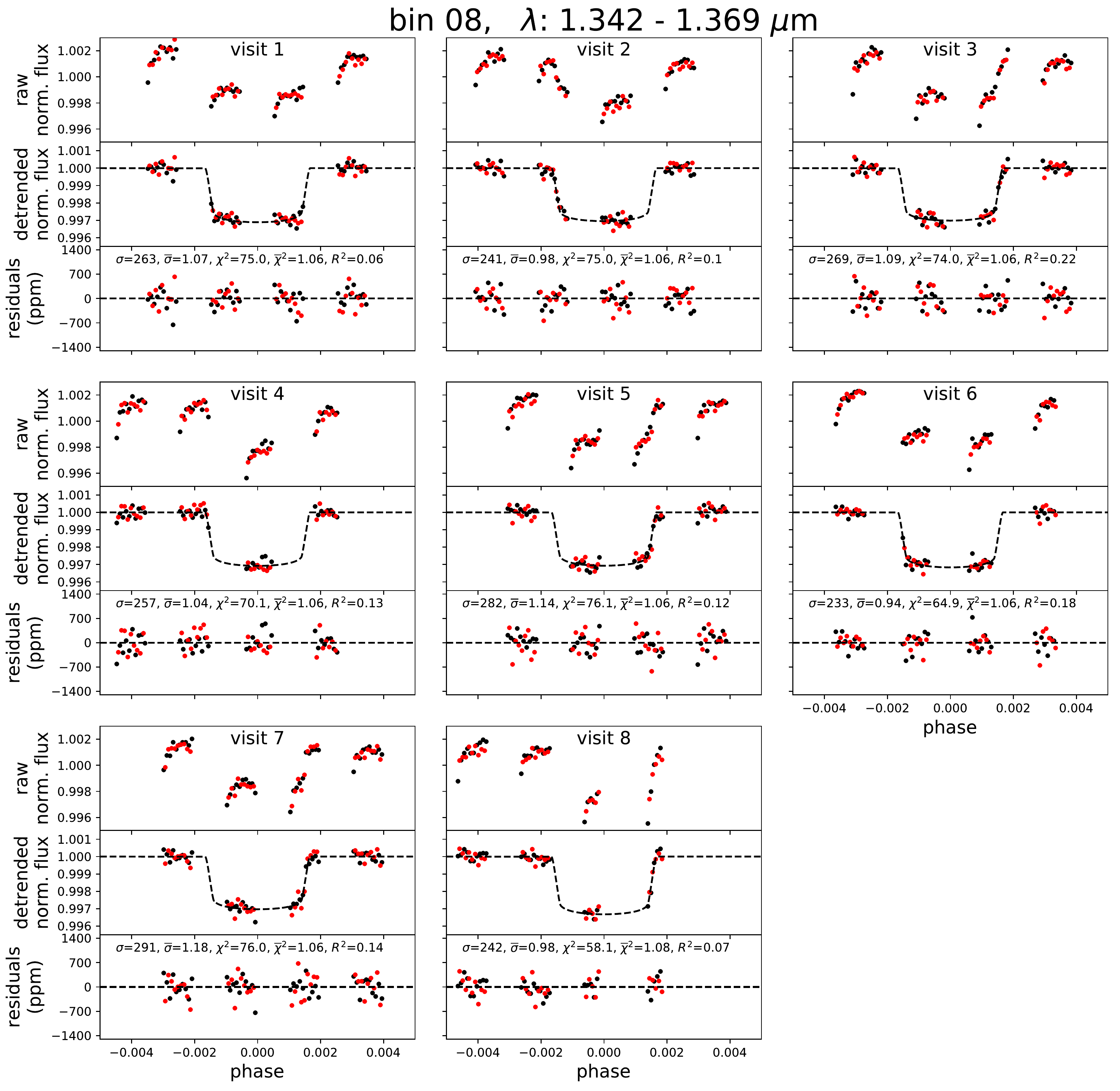}
\caption{Same as Supplementary Figure 2 for the spectral channel between 1.342 and 1.369 $\mu$m. }
\label{fig:bin_08_fit}
\end{figure}

\newpage
\begin{figure}
\centering
\includegraphics[width=\textwidth]{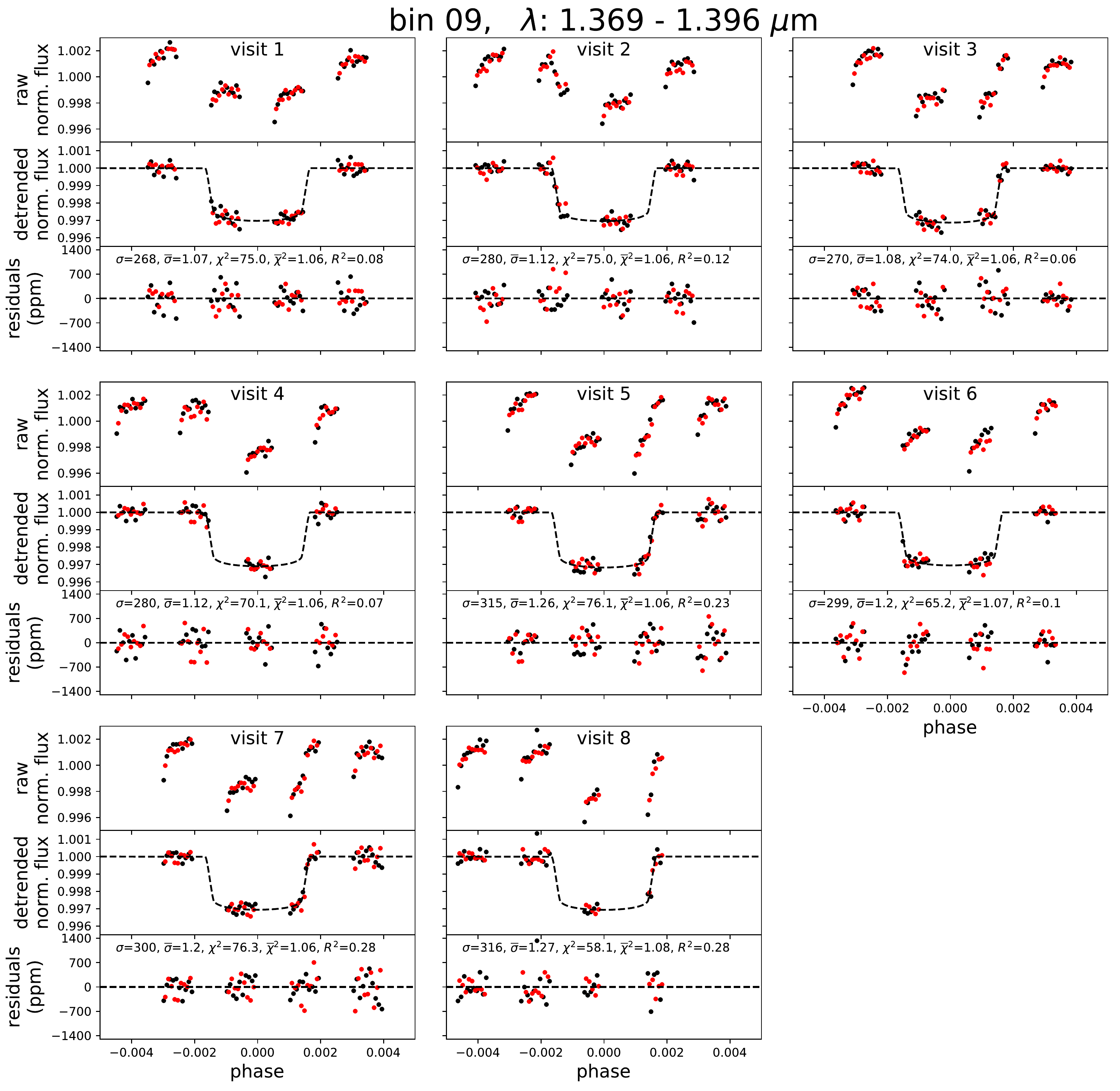}
\caption{Same as Supplementary Figure 2 for the spectral channel between 1.369 and 1.396 $\mu$m. }
\label{fig:bin_09_fit}
\end{figure}

\newpage
\begin{figure}
\centering
\includegraphics[width=\textwidth]{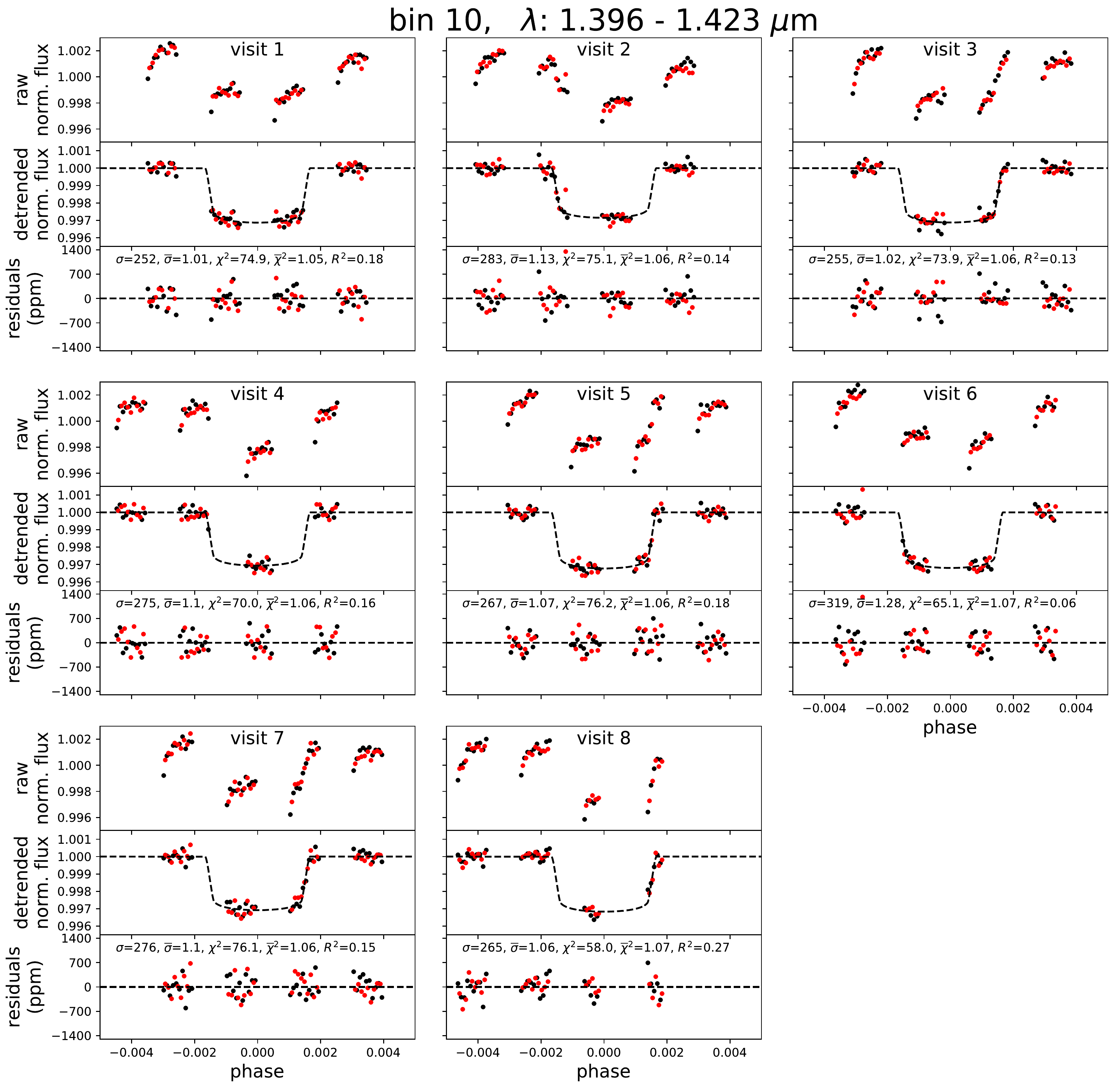}
\caption{Same as Supplementary Figure 2 for the spectral channel between 1.396 and 1.423 $\mu$m. }
\label{fig:bin_10_fit}
\end{figure}

\newpage
\begin{figure}
\centering
\includegraphics[width=\textwidth]{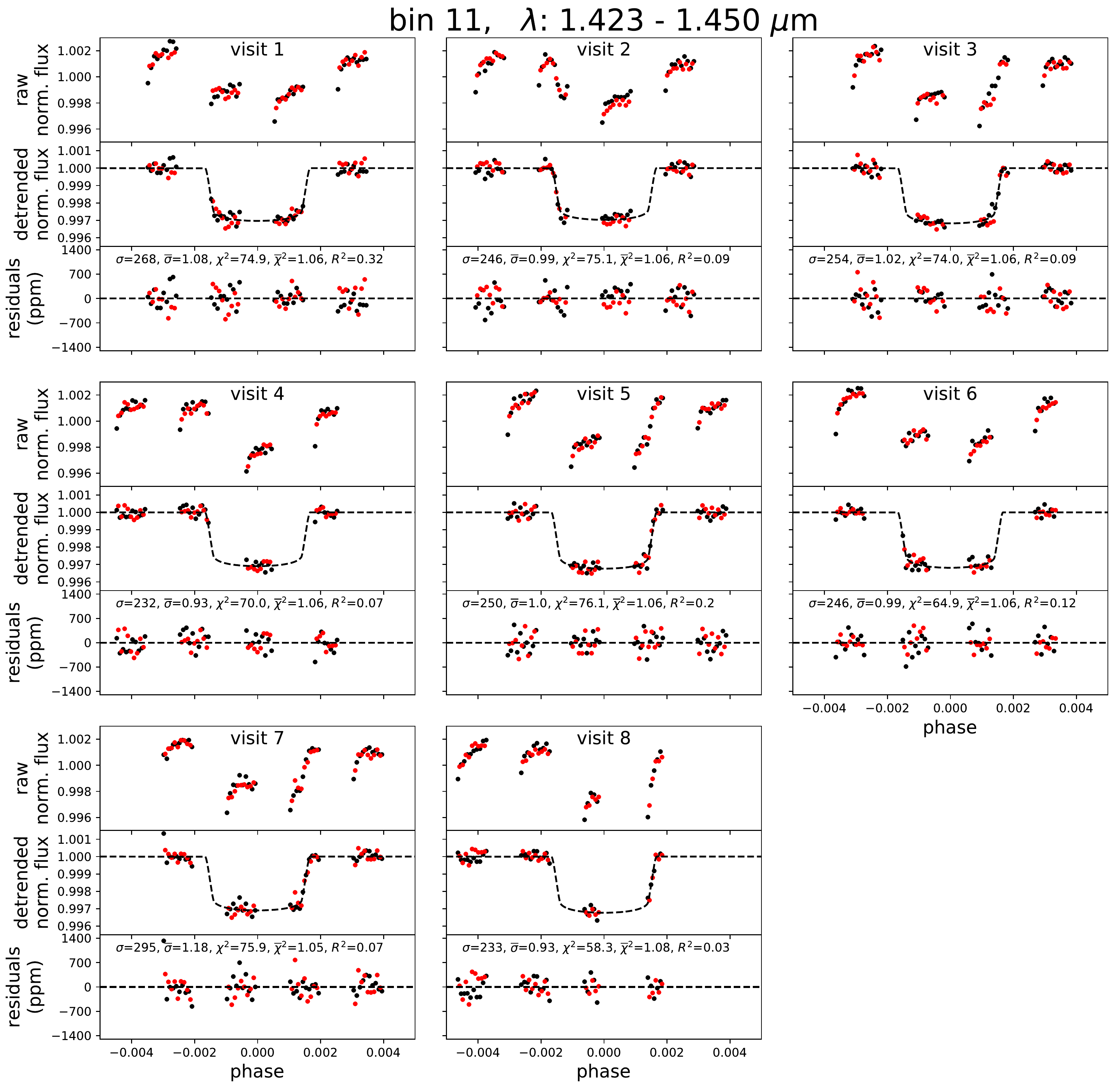}
\caption{Same as Supplementary Figure 2 for the spectral channel between 1.423 and 1.450 $\mu$m. }
\label{fig:bin_11_fit}
\end{figure}

\newpage
\begin{figure}
\centering
\includegraphics[width=\textwidth]{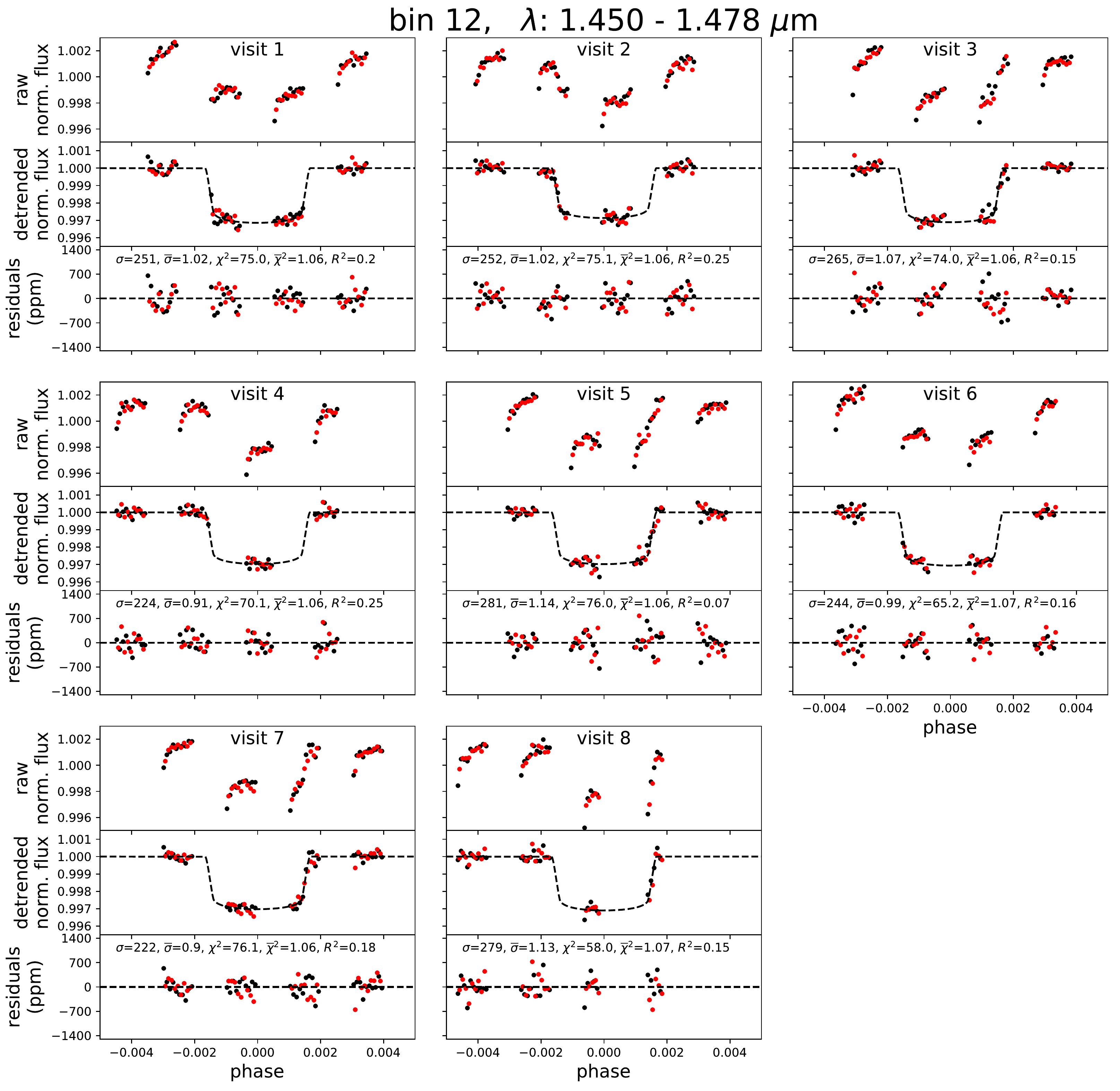}
\caption{Same as Supplementary Figure 2 for the spectral channel between 1.450 and 1.478 $\mu$m. }
\label{fig:bin_12_fit}
\end{figure}

\newpage
\begin{figure}
\centering
\includegraphics[width=\textwidth]{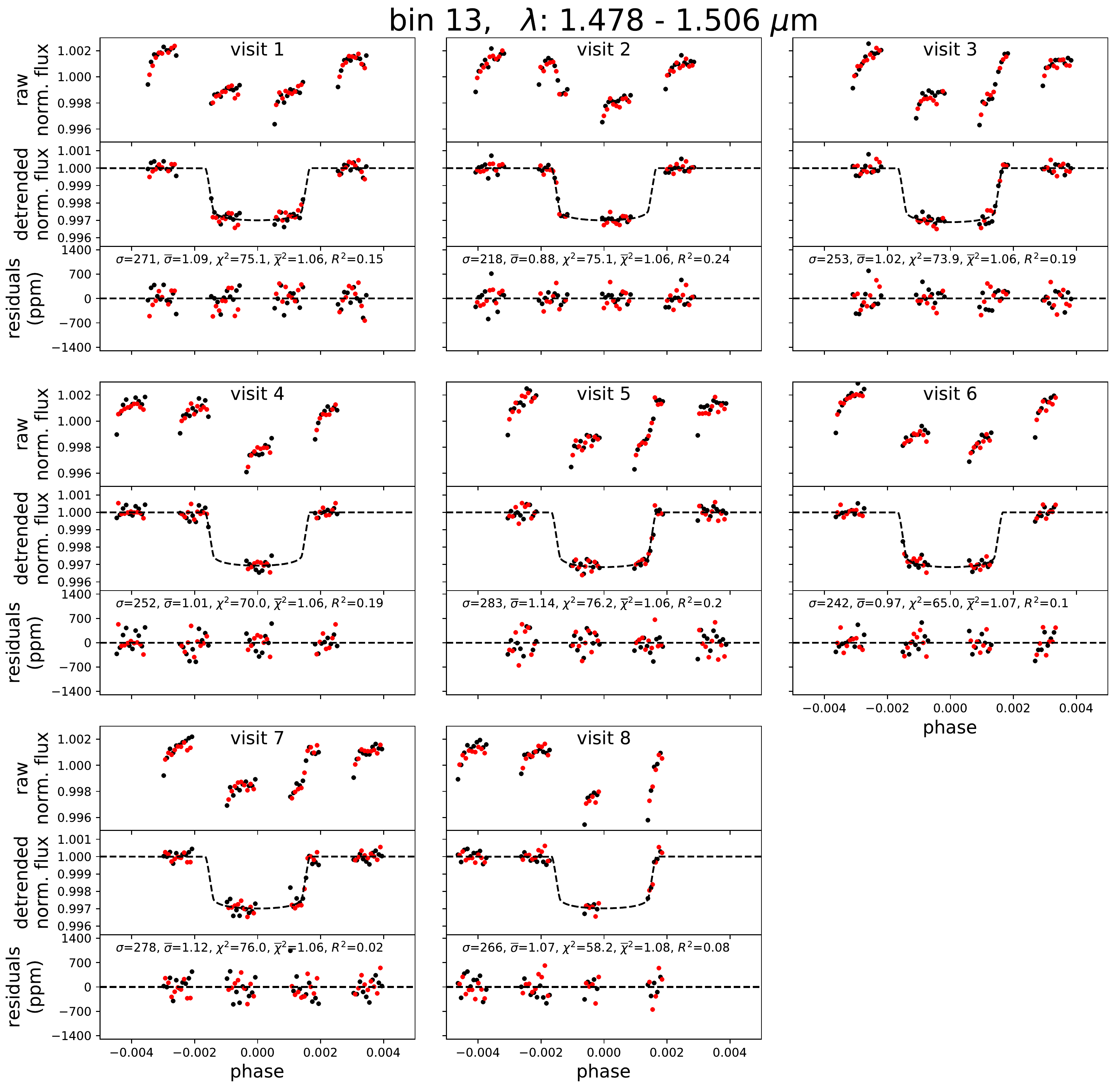}
\caption{Same as Supplementary Figure 2 for the spectral channel between 1.478 and 1.506 $\mu$m. }
\label{fig:bin_13_fit}
\end{figure}

\newpage
\begin{figure}
\centering
\includegraphics[width=\textwidth]{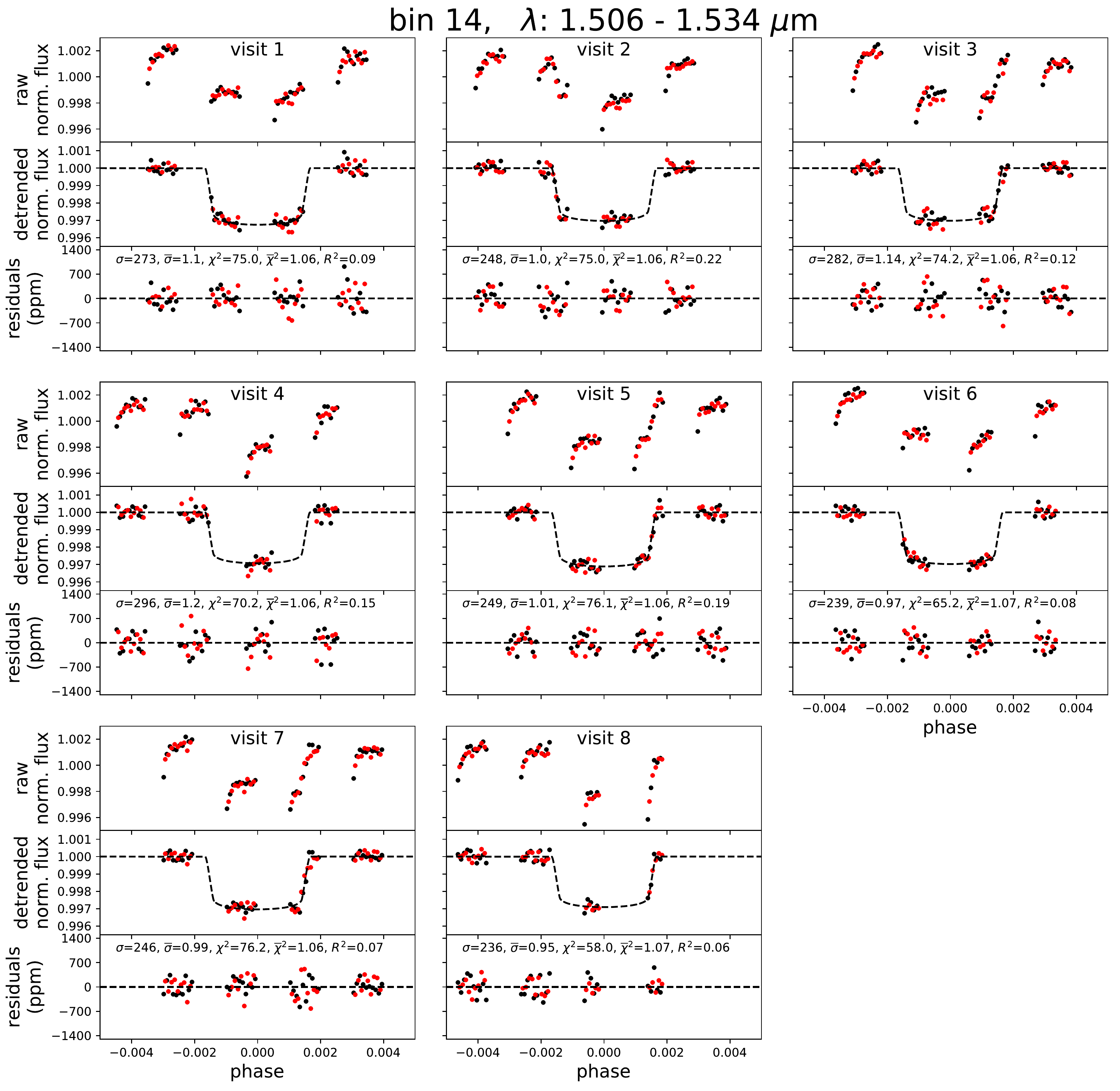}
\caption{Same as Supplementary Figure 2 for the spectral channel between 1.506 and 1.534 $\mu$m. }
\label{fig:bin_14_fit}
\end{figure}

\newpage
\begin{figure}
\centering
\includegraphics[width=\textwidth]{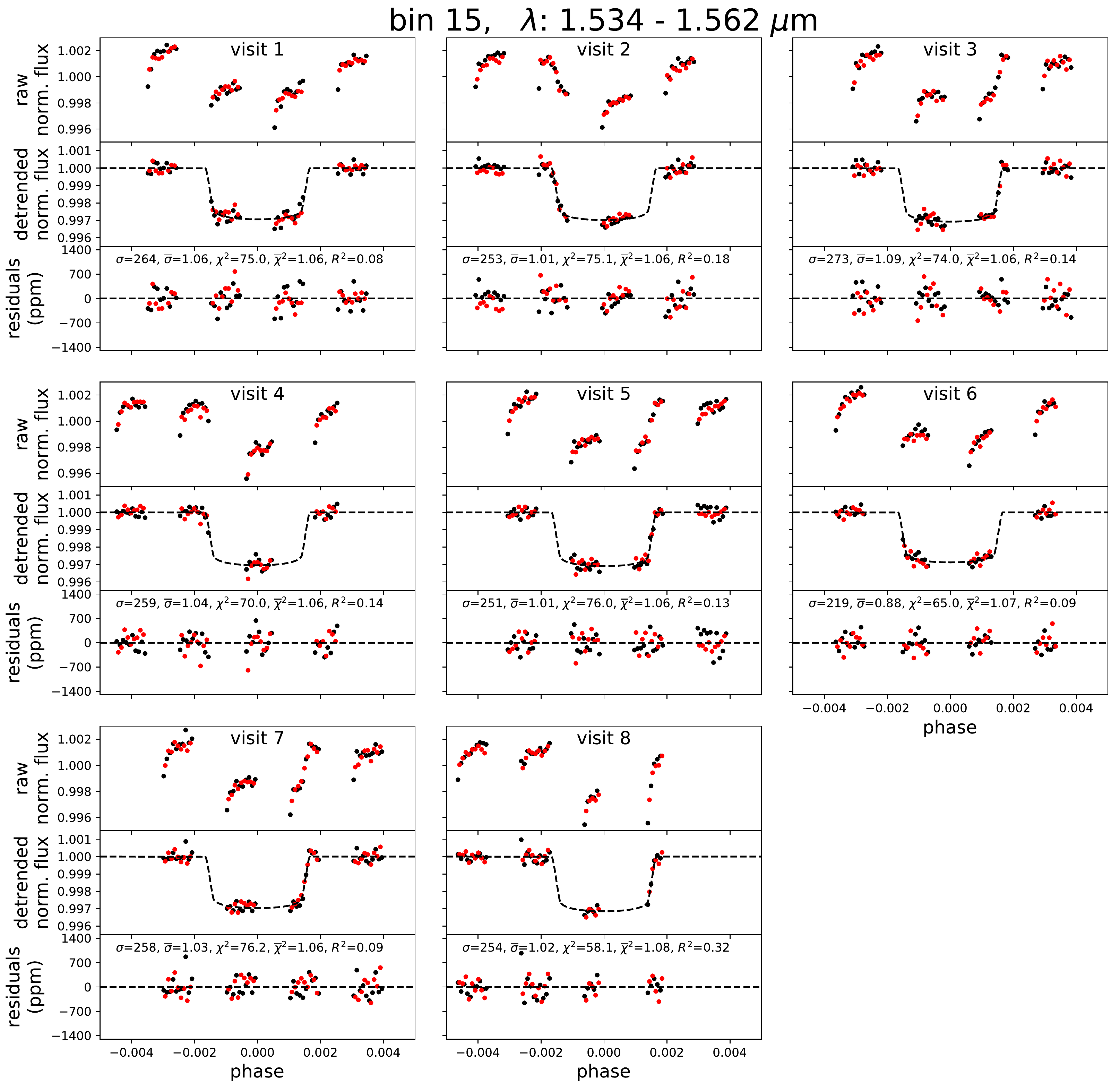}
\caption{Same as Supplementary Figure 2 for the spectral channel between 1.534 and 1.562 $\mu$m. }
\label{fig:bin_15_fit}
\end{figure}

\newpage
\begin{figure}
\centering
\includegraphics[width=\textwidth]{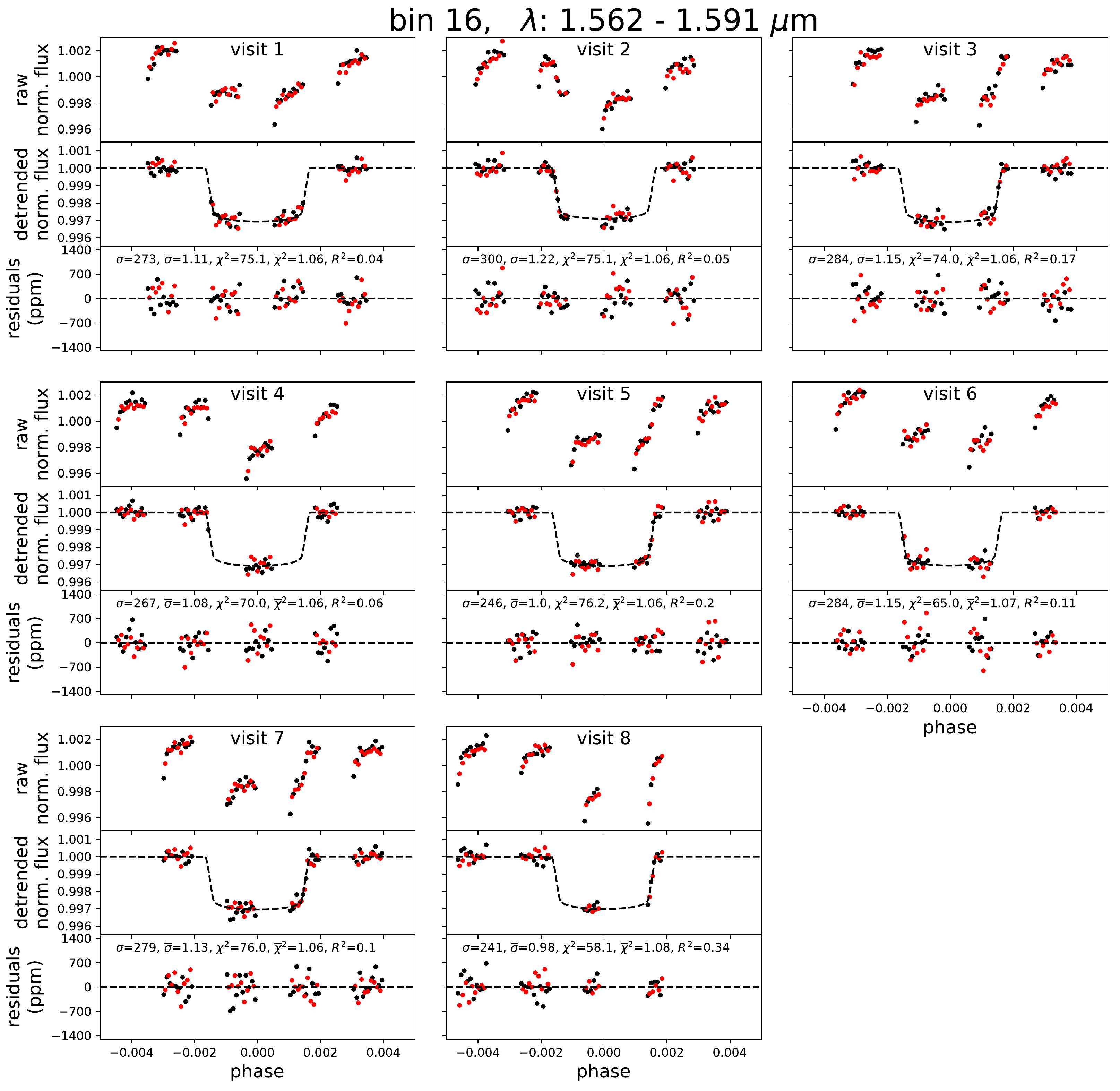}
\caption{Same as Supplementary Figure 2 for the spectral channel between 1.562 and 1.591 $\mu$m. }
\label{fig:bin_16_fit}
\end{figure}

\newpage
\begin{figure}
\centering
\includegraphics[width=\textwidth]{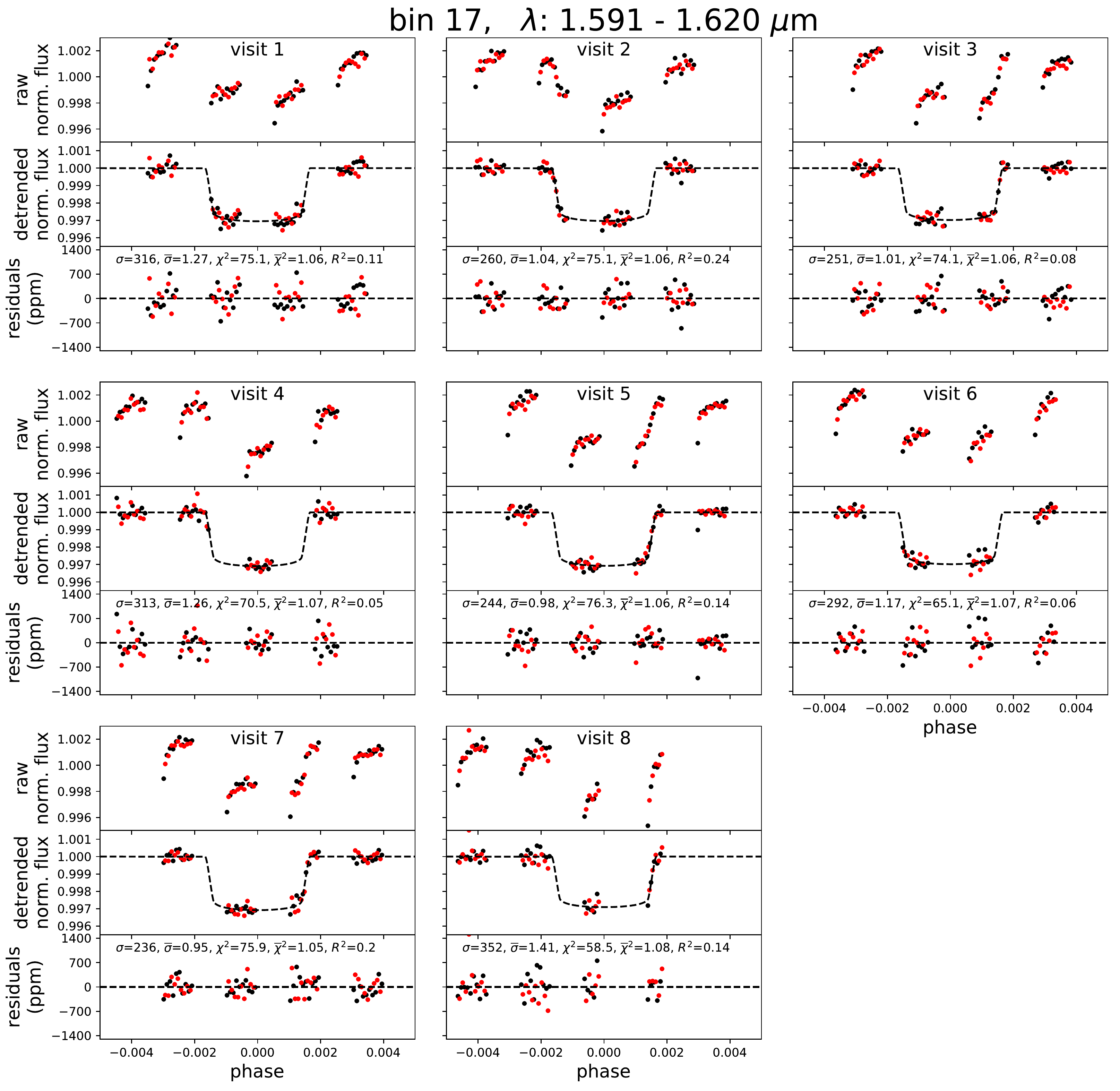}
\caption{Same as Supplementary Figure 2 for the spectral channel between 1.591 and 1.620 $\mu$m. }
\label{fig:bin_17_fit}
\end{figure}

\newpage
\begin{figure}
\centering
\includegraphics[width=\textwidth]{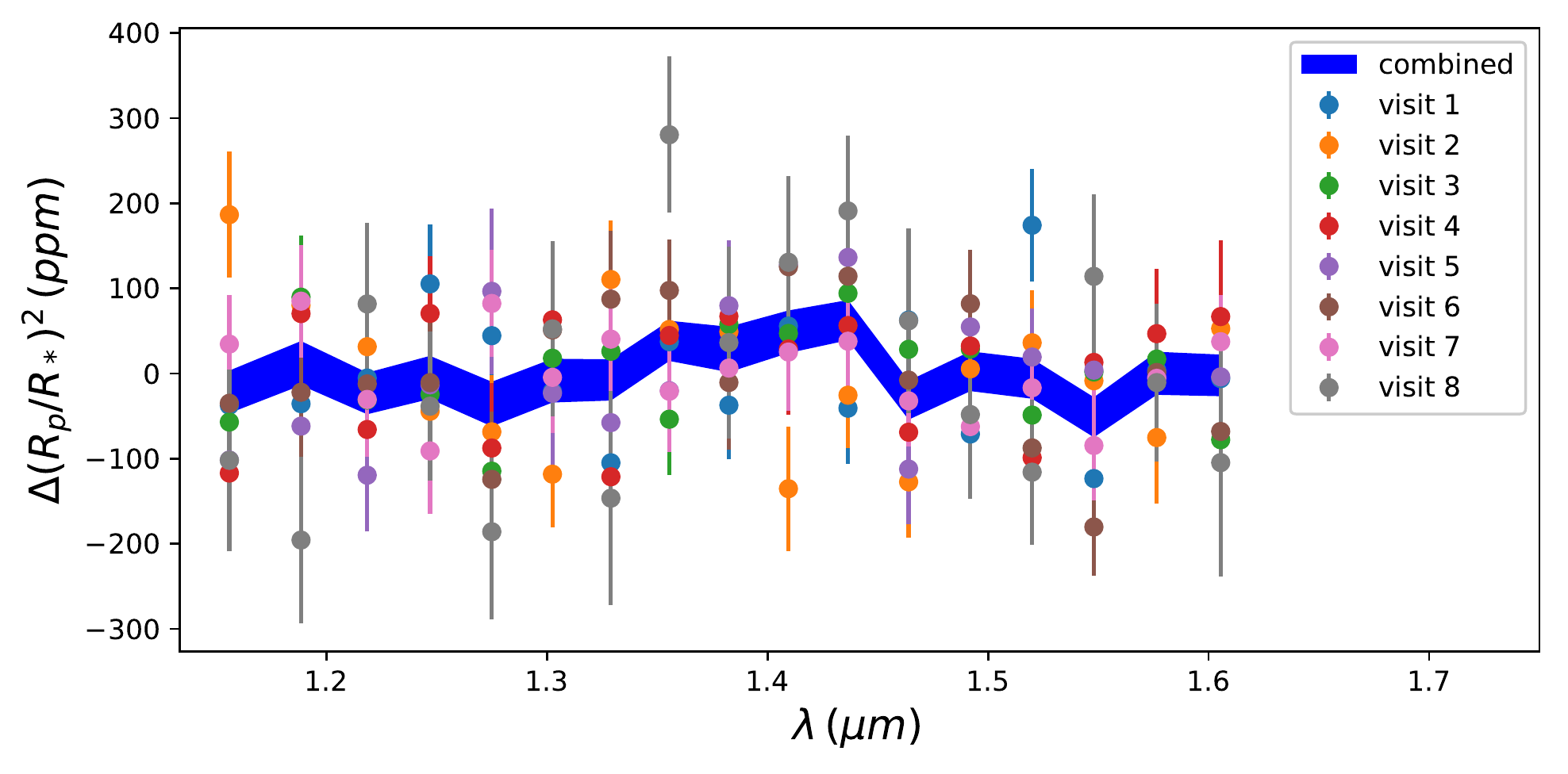}
\caption{Mean-subtracted spectra extracted per visits, and the final, reported, weighted average.}
\label{fig:all_spectra}
\end{figure}

\newpage
\begin{figure}
\centering
\includegraphics[width=\textwidth]{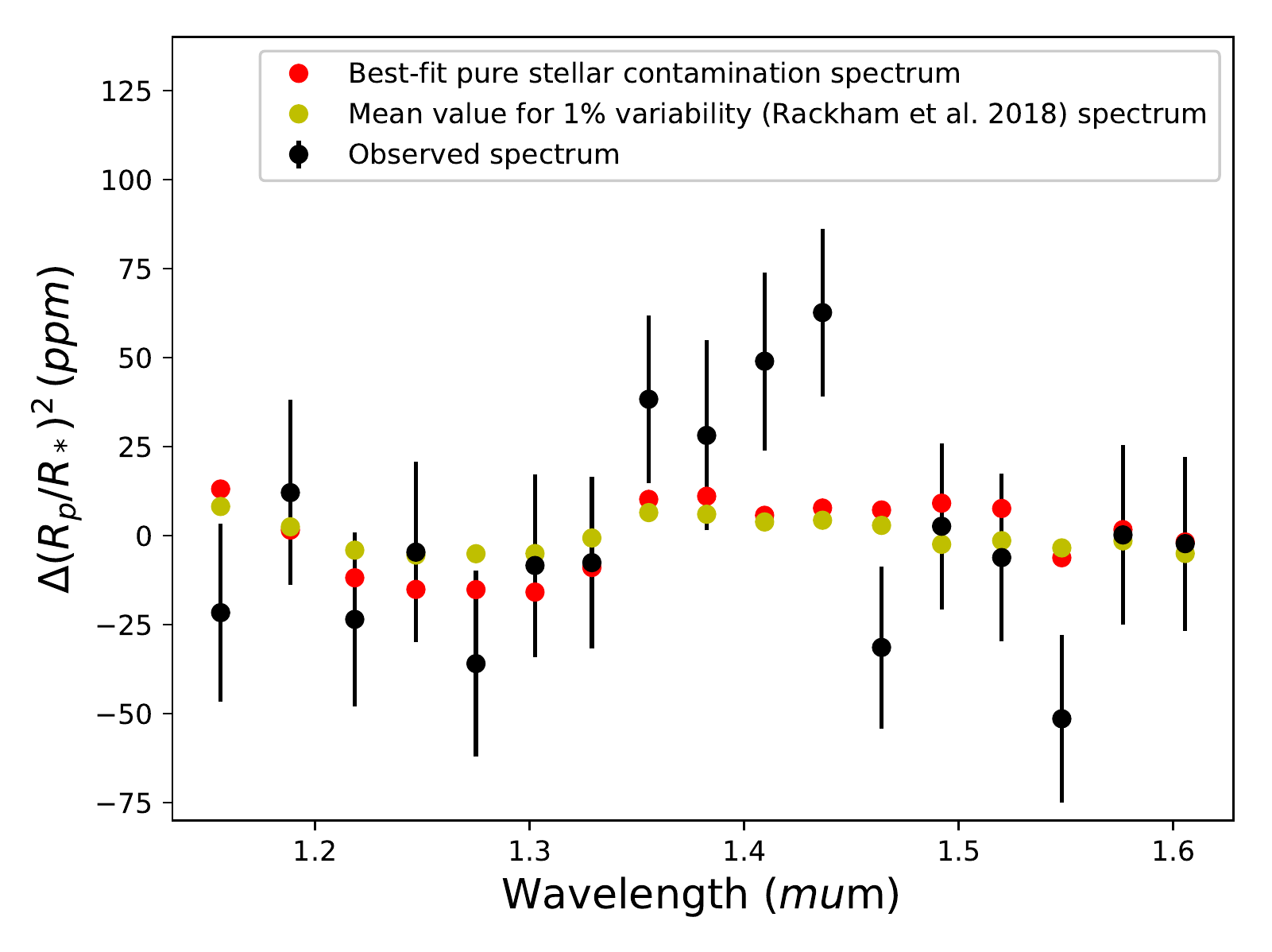}
\caption{Comparison between the observed spectrum and spectra produced only by the effect of spots and faculae (assuming no contribution from the planet). The yellow spectrum corresponds to the mean spot and faculae coverage predicted by Rackham et al. 2018\cite{Rackham2018} for an M2 star and 1\% I-band variability (19\%, 72\%, respectively). The red spectrum corresponds to the best-fit contamination model to the observed spectrum (26\%, 73\%, respectively). From this, we conclude that pure stellar contamination cannot explain the observed water feature.}
\label{fig:stellar_contamination}
\end{figure}


\begin{thebibliography}{10}
\expandafter\ifx\csname url\endcsname\relax
  \def\url#1{\texttt{#1}}\fi
\expandafter\ifx\csname urlprefix\endcsname\relax\def\urlprefix{URL }\fi
\providecommand{\bibinfo}[2]{#2}
\providecommand{\eprint}[2][]{\url{#2}}

\bibitem{Tinetti2007B2007ApJ...654L..99T}
\bibinfo{author}{{Tinetti}, G.} \emph{et~al.}
\newblock \bibinfo{title}{{Infrared Transmission Spectra for Extrasolar Giant
  Planets}}.
\newblock \emph{\bibinfo{journal}{\apjl}} \textbf{\bibinfo{volume}{654}},
  \bibinfo{pages}{L99--L102} (\bibinfo{year}{2007}).

\bibitem{Grillmair2008}
\bibinfo{author}{{Grillmair}, C.~J.} \emph{et~al.}
\newblock \bibinfo{title}{{Strong water absorption in the dayside emission spectrum of the planet HD189733b}}.
\newblock \emph{\bibinfo{journal}{Nature}} \textbf{\bibinfo{volume}{456}},
  \bibinfo{pages}{767-769} (\bibinfo{year}{2008}).
  
  \bibitem{Fraine2014}
\bibinfo{author}{{Fraine}, J.} \emph{et~al.}
\newblock \bibinfo{title}{{Water vapour absorption in the clear atmosphere of a Neptune-sized exoplanet}}.
\newblock \emph{\bibinfo{journal}{Nature}} \textbf{\bibinfo{volume}{350}},
  \bibinfo{pages}{64-67} (\bibinfo{year}{2015}).
  
  \bibitem{Macintosh2015}
\bibinfo{author}{{Macintosh}, B.} \emph{et~al.}
\newblock \bibinfo{title}{{Discovery and spectroscopy of the young jovian planet 51 Eri b with the Gemini Planet Imager}}.
\newblock \emph{\bibinfo{journal}{Science}} \textbf{\bibinfo{volume}{456}},
  \bibinfo{pages}{767-769} (\bibinfo{year}{2008}).
  
\bibitem{Tsiaras2018}
\bibinfo{author}{{Tsiaras}, A.} \emph{et~al.}
\newblock \bibinfo{title}{{A Population Study of Gaseous Exoplanets}}.
\newblock \emph{\bibinfo{journal}{\aj}} \textbf{\bibinfo{volume}{155}}
  (\bibinfo{year}{2018}).
  
\bibitem{deWit2018}
\bibinfo{author}{{de Wit}, J.} \emph{et~al.}
\newblock \bibinfo{title}{{Atmospheric reconnaissance of the habitable-zone
  Earth-sized planets orbiting TRAPPIST-1}}.
\newblock \emph{\bibinfo{journal}{Nat. Astron.}}
  \textbf{\bibinfo{volume}{2}}, \bibinfo{pages}{214--219}
  (\bibinfo{year}{2018})
  
\bibitem{Montet2015}
\bibinfo{author}{{Montet}, B.~T.} \emph{et~al.}
\newblock \bibinfo{title}{{Stellar and Planetary Properties of K2 Campaign 1
  Candidates and Validation of 17 Planets, Including a Planet Receiving
  Earth-like Insolation}}.
\newblock \emph{\bibinfo{journal}{\apj}} \textbf{\bibinfo{volume}{809}},
  \bibinfo{pages}{25} (\bibinfo{year}{2015}).

\bibitem{Benneke2013}
\bibinfo{author}{{Benneke}, B.} \& \bibinfo{author}{{Seager}, S.}
\newblock \bibinfo{title}{{How to Distinguish between Cloudy Mini-Neptunes and
  Water/Volatile-dominated Super-Earths}}.
\newblock \emph{\bibinfo{journal}{\apj}} \textbf{\bibinfo{volume}{778}},
  \bibinfo{pages}{153} (\bibinfo{year}{2013}).

\bibitem{Waldmann2015B2015ApJ...802..107W}
\bibinfo{author}{{Waldmann}, I.~P.} \emph{et~al.}
\newblock \bibinfo{title}{{Tau-REx I: A Next Generation Retrieval Code for
  Exoplanetary Atmospheres}}.
\newblock \emph{\bibinfo{journal}{\apj}} \textbf{\bibinfo{volume}{802}},
  \bibinfo{pages}{107} (\bibinfo{year}{2015}).

\bibitem{Segura2005}
\bibinfo{author}{{Segura}, A.} \emph{et~al.}
\newblock \bibinfo{title}{{Biosignatures from Earth-Like Planets Around M
  Dwarfs}}.
\newblock \emph{\bibinfo{journal}{Astrobiology}} \textbf{\bibinfo{volume}{5}},
  \bibinfo{pages}{706--725} (\bibinfo{year}{2005}).

\bibitem{Wordsworth2011}
\bibinfo{author}{{Wordsworth}, R.~D.} \emph{et~al.}
\newblock \bibinfo{title}{{Gliese 581d is the First Discovered Terrestrial-mass
  Exoplanet in the Habitable Zone}}.
\newblock \emph{\bibinfo{journal}{\apj}} \textbf{\bibinfo{volume}{733}},
  \bibinfo{pages}{L48} (\bibinfo{year}{2011}).

\bibitem{Leconte2013}
\bibinfo{author}{{Leconte}, J.} \emph{et~al.}
\newblock \bibinfo{title}{{3D climate modeling of close-in land planets:
  Circulation patterns, climate moist bistability, and habitability}}.
\newblock \emph{\bibinfo{journal}{\aap}} \textbf{\bibinfo{volume}{554}},
  \bibinfo{pages}{A69} (\bibinfo{year}{2013}).

\bibitem{Turbet2016}
\bibinfo{author}{{Turbet}, M.} \emph{et~al.}
\newblock \bibinfo{title}{{The habitability of Proxima Centauri b. II. Possible
  climates and observability}}.
\newblock \emph{\bibinfo{journal}{\aap}} \textbf{\bibinfo{volume}{596}},
  \bibinfo{pages}{A112} (\bibinfo{year}{2016}).

\bibitem{Deming2013}
\bibinfo{author}{{Deming}, D.} \emph{et~al.}
\newblock \bibinfo{title}{{Infrared Transmission Spectroscopy of the Exoplanets
  HD 209458b and XO-1b Using the Wide Field Camera-3 on the Hubble Space
  Telescope}}.
\newblock \emph{\bibinfo{journal}{\apj}} \textbf{\bibinfo{volume}{774}},
  \bibinfo{pages}{95} (\bibinfo{year}{2013}).

\bibitem{Kreidberg2014B2014Natur.505...69K}
\bibinfo{author}{{Kreidberg}, L.} \emph{et~al.}
\newblock \bibinfo{title}{{Clouds in the atmosphere of the super-Earth
  exoplanet GJ1214b}}.
\newblock \emph{\bibinfo{journal}{\nat}} \textbf{\bibinfo{volume}{505}},
  \bibinfo{pages}{69--72} (\bibinfo{year}{2014}).

\bibitem{Knutson2014B2014ApJ...794..155K}
\bibinfo{author}{{Knutson}, H.~A.} \emph{et~al.}
\newblock \bibinfo{title}{{Hubble Space Telescope Near-IR Transmission
  Spectroscopy of the Super-Earth HD 97658b}}.
\newblock \emph{\bibinfo{journal}{\apj}} \textbf{\bibinfo{volume}{794}},
  \bibinfo{pages}{155} (\bibinfo{year}{2014}).

\bibitem{Tsiaras2016B2016ApJ...820...99T}
\bibinfo{author}{{Tsiaras}, A.} \emph{et~al.}
\newblock \bibinfo{title}{{Detection of an Atmosphere Around the Super-Earth 55
  Cancri e}}.
\newblock \emph{\bibinfo{journal}{\apj}} \textbf{\bibinfo{volume}{820}},
  \bibinfo{pages}{99} (\bibinfo{year}{2016}).

\bibitem{Wakeford2019}
\bibinfo{author}{{Wakeford}, H.~R} \emph{et~al.}
\newblock \bibinfo{title}{{Disentangling the Planet from the Star in Late-Type M Dwarfs: A Case Study of TRAPPIST-1g}}.
\newblock \emph{\bibinfo{journal}{\aj}}
  \textbf{\bibinfo{volume}{157}}, \bibinfo{pages}{11}
  (\bibinfo{year}{2019}).

\bibitem{Benneke2017}
\bibinfo{author}{{Benneke}, B.} \emph{et~al.}
\newblock \bibinfo{title}{{Spitzer Observations Confirm and Rescue the
  Habitable-zone Super-Earth K2-18b for Future Characterization}}.
\newblock \emph{\bibinfo{journal}{\apj}} \textbf{\bibinfo{volume}{834}},
  \bibinfo{pages}{187} (\bibinfo{year}{2017}).

\bibitem{Kopparapu2013}
\bibinfo{author}{{Kopparapu}, R.~K.}
\newblock \bibinfo{title}{{A Revised Estimate of the Occurrence Rate of
  Terrestrial Planets in the Habitable Zones around Kepler M-dwarfs}}.
\newblock \emph{\bibinfo{journal}{\apjl}} \textbf{\bibinfo{volume}{767}},
  \bibinfo{pages}{L8} (\bibinfo{year}{2013}).

\bibitem{Valencia2018}
\bibinfo{author}{{Valencia}, D.}, \bibinfo{author}{{Tan}, V. Y.~Y.} \&
  \bibinfo{author}{{Zajac}, Z.}
\newblock \bibinfo{title}{{Habitability from Tidally Induced Tectonics}}.
\newblock \emph{\bibinfo{journal}{\apj}} \textbf{\bibinfo{volume}{857}},
  \bibinfo{pages}{106} (\bibinfo{year}{2018}).

\bibitem{Cloutier2017}
\bibinfo{author}{{Cloutier}, R.} \emph{et~al.}
\newblock \bibinfo{title}{{Characterization of the K2-18 multi-planetary system
  with HARPS. A habitable zone super-Earth and discovery of a second, warm
  super-Earth on a non-coplanar orbit}}.
\newblock \emph{\bibinfo{journal}{\aap}} \textbf{\bibinfo{volume}{608}},
  \bibinfo{pages}{A35} (\bibinfo{year}{2017}).

\bibitem{Valencia2013}
\bibinfo{author}{{Valencia}, D.}, \bibinfo{author}{{Guillot}, T.},
  \bibinfo{author}{{Parmentier}, V.} \& \bibinfo{author}{{Freedman}, R.~S.}
\newblock \bibinfo{title}{{Bulk Composition of GJ 1214b and Other Sub-Neptune
  Exoplanets}}.
\newblock \emph{\bibinfo{journal}{\apj}} \textbf{\bibinfo{volume}{775}},
  \bibinfo{pages}{10} (\bibinfo{year}{2013}).

\bibitem{Zeng2016}
\bibinfo{author}{{Zeng}, L.}, \bibinfo{author}{{Sasselov}, D.~D.} \&
  \bibinfo{author}{{Jacobsen}, S.~B.}
\newblock \bibinfo{title}{{Mass-Radius Relation for Rocky Planets Based on
  PREM}}.
\newblock \emph{\bibinfo{journal}{\apj}} \textbf{\bibinfo{volume}{819}},
  \bibinfo{pages}{127} (\bibinfo{year}{2016}).
  
\bibitem{Tsiaras2016B2016ApJ...832..202T}
\bibinfo{author}{{Tsiaras}, A.} \emph{et~al.}
\newblock \bibinfo{title}{{A New Approach to Analyzing HST Spatial Scans: The
  Transmission Spectrum of HD 209458 b}}.
\newblock \emph{\bibinfo{journal}{\apj}} \textbf{\bibinfo{volume}{832}},
  \bibinfo{pages}{202} (\bibinfo{year}{2016}).

\bibitem{Eastman2010}
\bibinfo{author}{{Eastman}, J.} \emph{et~al.}
\newblock \bibinfo{title}{{Achieving Better Than 1 Minute Accuracy in the Heliocentric and Barycentric Julian Dates}}.
\newblock \emph{\bibinfo{journal}{\pasp}} \textbf{\bibinfo{volume}{122}},
  \bibinfo{pages}{935} (\bibinfo{year}{2010}).

\bibitem{Waldmann2015B2015ApJ...813...13W}
\bibinfo{author}{{Waldmann}, I.~P.} \emph{et~al.}
\newblock \bibinfo{title}{{Tau-REx II: Retrieval of Emission Spectra}}.
\newblock \emph{\bibinfo{journal}{\apj}} \textbf{\bibinfo{volume}{813}},
  \bibinfo{pages}{13} (\bibinfo{year}{2015}).

\bibitem{Tennyson2016}
\bibinfo{author}{{Tennyson}, J.} \emph{et~al.}
\newblock \bibinfo{title}{{The ExoMol database: Molecular line lists for
  exoplanet and other hot atmospheres}}.
\newblock \emph{\bibinfo{journal}{J. Mol. Spec.}}
  \textbf{\bibinfo{volume}{327}}, \bibinfo{pages}{73--94}
  (\bibinfo{year}{2016}).

\bibitem{Tinetti2018}
\bibinfo{author}{{Tinetti}, G.} \emph{et~al.}
\newblock \bibinfo{title}{{A chemical survey of exoplanets with ARIEL}}.
\newblock \emph{\bibinfo{journal}{Exp. Astron.}}
  \textbf{\bibinfo{volume}{46}}, \bibinfo{pages}{135--209}
  (\bibinfo{year}{2018}).

\end{thebibliography}

\begin{thebibliography}{10}
\expandafter\ifx\csname url\endcsname\relax
  \def\url#1{\texttt{#1}}\fi
\expandafter\ifx\csname urlprefix\endcsname\relax\def\urlprefix{URL }\fi
\providecommand{\bibinfo}[2]{#2}
\providecommand{\eprint}[2][]{\url{#2}}

\makeatletter
\addtocounter{\@listctr}{30}
\makeatother

\bibitem{Allard2012}
\bibinfo{author}{{Allard}, F.}, \bibinfo{author}{{Homeier}, D.} \&
  \bibinfo{author}{{Freytag}, B.}
\newblock \bibinfo{title}{{Models of very-low-mass stars, brown dwarfs and
  exoplanets}}.
\newblock \emph{\bibinfo{journal}{Philos. Trans. Royal
  Soc. A}} \textbf{\bibinfo{volume}{370}},
  \bibinfo{pages}{2765--2777} (\bibinfo{year}{2012}).

\bibitem{Claret2000}
\bibinfo{author}{{Claret}, A.}
\newblock \bibinfo{title}{{A new non-linear limb-darkening law for LTE stellar
  atmosphere models. Calculations for -5.0 $<$= log[M/H] $<$= +1, 2000 K $<$=
  T$_{eff}$ $<$= 50000 K at several surface gravities}}.
\newblock \emph{\bibinfo{journal}{\aap}} \textbf{\bibinfo{volume}{363}},
  \bibinfo{pages}{1081--1190} (\bibinfo{year}{2000}).

\bibitem{Kreidberg2015}
\bibinfo{author}{{Kreidberg}, L.} \emph{et~al.}
\newblock \bibinfo{title}{{A Detection of Water in the Transmission Spectrum of
  the Hot Jupiter WASP-12b and Implications for Its Atmospheric Composition}}.
\newblock \emph{\bibinfo{journal}{\apj}} \textbf{\bibinfo{volume}{814}},
  \bibinfo{pages}{66} (\bibinfo{year}{2015}).

\bibitem{Evans2016}
\bibinfo{author}{{Evans}, T.~M.} \emph{et~al.}
\newblock \bibinfo{title}{{Detection of H$_{2}$O and Evidence for TiO/VO in an
  Ultra-hot Exoplanet Atmosphere}}.
\newblock \emph{\bibinfo{journal}{\apjl}} \textbf{\bibinfo{volume}{822}},
  \bibinfo{pages}{L4} (\bibinfo{year}{2016}).

\bibitem{Line2016B2016AJ....152..203L}
\bibinfo{author}{{Line}, M.~R.} \emph{et~al.}
\newblock \bibinfo{title}{{No Thermal Inversion and a Solar Water Abundance for
  the Hot Jupiter HD 209458b from HST/WFC3 Spectroscopy}}.
\newblock \emph{\bibinfo{journal}{\aj}} \textbf{\bibinfo{volume}{152}},
  \bibinfo{pages}{203} (\bibinfo{year}{2016}).

\bibitem{Wakeford2017B2017ApJ...835L..12W}
\bibinfo{author}{{Wakeford}, H.~R.} \emph{et~al.}
\newblock \bibinfo{title}{{HST PanCET program: A Cloudy Atmosphere for the
  Promising JWST Target WASP-101b}}.
\newblock \emph{\bibinfo{journal}{\apjl}} \textbf{\bibinfo{volume}{835}},
  \bibinfo{pages}{L12} (\bibinfo{year}{2017}).

\bibitem{McCullough2012}
\bibinfo{author}{{McCullough}, P.} \& \bibinfo{author}{{MacKenty}, J.}
\newblock \bibinfo{title}{{Considerations for using Spatial Scans with WFC3}}.
\newblock \bibinfo{type}{Tech. Rep.} (\bibinfo{year}{2012}).

\bibitem{Skilling2006}
\bibinfo{author}{{Skilling}, J.}
\newblock \bibinfo{title}{{Nested sampling for general Bayesian computation}}.
\newblock \emph{\bibinfo{journal}{Bayesian Analysis}}
  \textbf{\bibinfo{volume}{1}}, \bibinfo{pages}{833--860}
  (\bibinfo{year}{2006}).

\bibitem{Feroz2009}
\bibinfo{author}{{Feroz}, F.}, \bibinfo{author}{{Hobson}, M.~P.} \&
  \bibinfo{author}{{Bridges}, M.}
\newblock \bibinfo{title}{{MULTINEST: an efficient and robust Bayesian
  inference tool for cosmology and particle physics}}.
\newblock \emph{\bibinfo{journal}{\mnras}} \textbf{\bibinfo{volume}{398}},
  \bibinfo{pages}{1601--1614} (\bibinfo{year}{2009}).

\bibitem{Sarkis2018}
\bibinfo{author}{{Sarkis}, P.} \emph{et~al.}
\newblock \bibinfo{title}{{The CARMENES Search for Exoplanets around M Dwarfs: A Low-mass Planet in the Temperate Zone of the Nearby K2-18}}.
\newblock \emph{\bibinfo{journal}{\aj}} \textbf{\bibinfo{volume}{155}},
  \bibinfo{pages}{257} (\bibinfo{year}{2018}).

\bibitem{Rackham2018}
\bibinfo{author}{{Rackham}, B.~V.}, \bibinfo{author}{{Apai}, D.} \& \bibinfo{author}{{Giampapa}, M.~S.}
\newblock \bibinfo{title}{{The Transit Light Source Effect: False Spectral Features and Incorrect Densities for M-dwarf Transiting Planets}}.
\newblock \emph{\bibinfo{journal}{\apj}} \textbf{\bibinfo{volume}{853}},
  \bibinfo{pages}{122} (\bibinfo{year}{2018}).

\bibitem{Barber2006}
\bibinfo{author}{{Barber}, R.~J.}, \bibinfo{author}{{Tennyson}, J.},
  \bibinfo{author}{{Harris}, G.~J.} \& \bibinfo{author}{{Tolchenov}, R.~N.}
\newblock \bibinfo{title}{{A high-accuracy computed water line list}}.
\newblock \emph{\bibinfo{journal}{\mnras}} \textbf{\bibinfo{volume}{368}},
  \bibinfo{pages}{1087--1094} (\bibinfo{year}{2006}).

\bibitem{Rothman2010}
\bibinfo{author}{{Rothman}, L.~S.} \emph{et~al.}
\newblock \bibinfo{title}{{HITEMP, the high-temperature molecular spectroscopic
  database}}.
\newblock \emph{\bibinfo{journal}{\jqsrt}} \textbf{\bibinfo{volume}{111}},
  \bibinfo{pages}{2139--2150} (\bibinfo{year}{2010}).

\bibitem{Yurchenko2014}
\bibinfo{author}{{Yurchenko}, S.~N.} \& \bibinfo{author}{{Tennyson}, J.}
\newblock \bibinfo{title}{{ExoMol line lists - IV. The rotation-vibration
  spectrum of methane up to 1500 K}}.
\newblock \emph{\bibinfo{journal}{\mnras}} \textbf{\bibinfo{volume}{440}},
  \bibinfo{pages}{1649--1661} (\bibinfo{year}{2014}).

\bibitem{Yurchenko2011}
\bibinfo{author}{{Yurchenko}, S.~N.}, \bibinfo{author}{{Barber}, R.~J.} \&
  \bibinfo{author}{{Tennyson}, J.}
\newblock \bibinfo{title}{{A variationally computed line list for hot
  NH$_{3}$}}.
\newblock \emph{\bibinfo{journal}{\mnras}} \textbf{\bibinfo{volume}{413}},
  \bibinfo{pages}{1828--1834} (\bibinfo{year}{2011}).

\bibitem{corner}
\bibinfo{author}{Foreman-Mackey, D.}
\newblock \bibinfo{title}{corner.py: Scatterplot matrices in python}.
\newblock \emph{\bibinfo{journal}{The Journal of Open Source Software}}
  \textbf{\bibinfo{volume}{24}} (\bibinfo{year}{2016}).

\end{thebibliography}
\end{document}